\documentclass[12pt]{article}
\pdfoutput=1
\usepackage[a4paper]{geometry}
\usepackage{jheppub, amsmath,amssymb,amsfonts,amsxtra, mathrsfs, makeidx,graphics,graphicx,amsthm,epsfig, bm,longtable,float, color,tikz,mathtools,xfrac,footnote,rotating, lscape, makecell, environ,mathtools, empheq}

\usepackage{subfig}
\usepackage{multirow}
\usepackage{adjustbox}

\usepackage{amsthm}
\usepackage{hyperref}

\usepackage{amsmath}
\usepackage{amsfonts}
\usepackage{commath}
\usepackage[english]{babel}
\usepackage{setspace}

\usepackage{epsfig,amssymb} 
\usepackage{amsmath}

\usepackage{amscd,color}
\usepackage{amsmath,graphicx}
\usepackage{verbatim,booktabs}

\usepackage{latexsym}
\usepackage{amsmath,amsfonts,amssymb,amsthm}
\usepackage{amsmath,amsthm}

\DeclareMathOperator{\Tr}{Tr}

\newcommand{\nn}{n}

\usepackage{tikz}
\usetikzlibrary{tikzmark}
\usetikzlibrary{math} 
\usetikzlibrary{matrix}
\usetikzlibrary{calc}
\usetikzlibrary{positioning}
\definecolor{azzurro}{RGB}{0,118,186}
\definecolor{rosso}{RGB}{255,0,0}

\usepackage{array}

\definecolor{mandarancio}{RGB}{255,147,0}
\definecolor{azzurro}{RGB}{0,118,186}
\definecolor{terrabruciatadiconcorezzo}{RGB}{181,23,0}

\def\centerarc[#1](#2)(#3:#4:#5)
{ \draw[#1] ($(#2)+({#5*cos(#3)},{#5*sin(#3)})$) arc (#3:#4:#5) }

\title{
\begin{center}
Webs of 3d $\mathcal{N}=2$ dualities \\ with D-type superpotentials  
\end{center}
}

\author[a]{Antonio Amariti}
\author[a,b]{and Simone 
 Rota}
\affiliation[a]{INFN, Sezione di Milano, Via Celoria 16, I-20133 Milano, Italy}

\affiliation[b]{Dipartimento di Fisica, Universit\`a degli Studi di Milano, Via Celoria 16, I-20133 Milano, Italy}
\emailAdd{antonio.amariti@mi.infn.it,  simone.rota@mi.infn.it}

\abstract{
We study 3d $\mathcal{N}=2$ dualities arising from the compactification of 4d $\mathcal{N}=1$ 
$Usp(2\nn)$ SQCD with two antisymmetric rank-two tensors and $D_{k+2}$-type superpotential, with odd $k$.
The analysis is carried out  by using field theory methods and by checking the various steps 
on the three sphere partition function. Most of the results are based on a conjectural confining duality that we do not
prove but that fits consistently with the web of dualities that we obtain.
Along the analysis we recover dualities already claimed in the literature and we propose new ones.
 The final picture that emerges fits with the general scheme  worked out for ordinary SQCD 
 and for adjoint SQCD with $A_k$-type superpotentials.}

\begin{document}

\maketitle

%
%
%
%
\section{Introduction}
\label{sec:intro}
%
%
%
%

Constructing 3d $\mathcal{N}=2$ dualities from the dimensional reduction of 4d $\mathcal{N}=1$ parents on $S^1$ has been a fertile field of research in the last decade (see \cite{Dolan:2011rp,Niarchos:2012ah,Aharony:2013dha,Aharony:2013kma,Csaki:2014cwa,Nii:2014jsa,Amariti:2014iza,Amariti:2015yea,Amariti:2015mva,Amariti:2015kha,Csaki:2017cqm,Benvenuti:2017kud,Benini:2017dud,Benvenuti:2017bpg,Nieri:2018pev,Hwang:2018uyj,Amariti:2018wht,Benvenuti:2018bav} for a partial list of references).
Furthermore the possibility of engineering real mass flows in 3d has been crucial for constructing large webs of dualities
without a 4d counterpart.
The two main reasons behind this behavior are the possibility of generating CS terms in the  action and 
the presence of a Coulomb branch associated to the monopole operators (we refer the reader to \cite{Aharony:1997bx,Intriligator:2013lca} for quite comprehensive reviews on 3d $\mathcal{N}=2$ gauge theories).

This 4d/3d dimensional reduction program has naturally started from Seiberg duality for SQCD \cite{Seiberg:1994pq} with unitary gauge group and it has been extended in various directions, either considering 4d dualities with real gauge groups \cite{Intriligator:1995id,Intriligator:1995ne}, or a rank-two tensor matter field with a power law superpotential \cite{Kutasov:1995ss} or both \cite{Leigh:1995qp,Intriligator:1995ax}.

Cases with two rank-two tensors are less studied \cite{Brodie:1996vx,Brodie:1996xm}, and only quite recently the case of $U(\nn)$ SQCD has been investigated \cite{Hwang:2018uyj}.
The 4d duality shows indeed some differences with respect to the cases with a single or without any adjoint. 
For example the chiral ring is conjectured to be truncated at quantum level in some cases and this duality does not 
have a brane engineering in the Hanany-Witten setup \cite{Hanany:1996ie} so far.

The analysis of \cite{Hwang:2018uyj} was based on the general 4d/3d prescription of \cite{Aharony:2013dha}:
the authors obtained first an effective duality on $S^1$ and then they arrived to the pure 3d limit by a real mass flow. Similarly to the analysis performed in the case with a single adjoint in \cite{Nii:2014jsa} the construction was pursued by breaking the $U(\nn)$ gauge group in a product of $U(n_i)$ SQCD sectors and then using known results from the reduction of 4d Seiberg duality to 3d $U(\nn)$  Aharony duality \cite{Aharony:1997gp}.
Eventually the unbroken gauge theory was reconstructed and the 3d duality was proposed.
Then in a more recent paper \cite{Hwang:2022jjs} a generalization of this duality, 
adding linear superpotentials for the  monopole and the anti-monopole with topological charge one,
 has been claimed, generalizing the ordinary SQCD construction of \cite{Benini:2017dud} 
and the case with an adjoint of \cite{Amariti:2018wht}.
In the case with two adjoints it was observed that there is a second type of superpotential, for the monopole and the anti-monopole with topological charge two.
Non trivial checks have been done by comparing  the expansion of the 3d superconformal index of 
\cite{Kim:2009wb}  up to  very high orders in the fugacities.
This is the state of the art in the analysis of 3d $\mathcal{N}=2$ SQCD with two adjoints. 
The large web of dualities obtained for other families of SQCD-like models has not been studied yet and further 
generalizations in this direction are desirable.

Motivated by this discussion in this paper we study 3d $\mathcal{N}=2$ $U(\nn)$ SQCD with two adjoints and 
$USp(2\nn)$ SQCD with two rank-two antisymmetric tensors, recovering some of the results already found in the literature and finding new dualities.

We start our analysis with the case of 4d $USp(2\nn)$ SQCD with $2N_f$ fundamentals and two rank-two antisymmetric tensors $A$ and $B$ and superpotential $W=AB^2+A^{k+1}$. This model was studied in \cite{Brodie:1996xm} and it was shown to be dual to $USp(2(3kN_f-\nn-4k-2))$ SQCD, with a similar matter content and a non trivial superpotential involving the dressed symmetric and anti-symmetric mesons of the electric model.
This model is the starting point of our analysis because it leads to the 3d dualities discussed in \cite{Hwang:2018uyj} and \cite{Hwang:2022jjs}, through dimensional reduction and real mass flows.
By circle reduction we first obtain an effective duality, with a KK monopole superpotential turned on. 
 Then through a real mass and a Higgs flow on both the electric and the magnetic side
we arrive at the 3d $U(\nn)$ duality recently discovered in \cite{Hwang:2022jjs} with two linear superpotential deformations, for the bare monopole $V_+$ and anti-monopole $V_-$ respectively. 

The next step consists of eliminating the monopole deformations. This is the more delicate and intricate step of our construction, because it requires a dual Higgsing of the gauge group. This dual Higgsing produces extra gauge sectors
that are expected to be dual to singlets. We conjecture the existence of such confining dualities, involving   
$U(k-1)$ SQCD with one flavor, two adjoints $X$ and $Y$ and superpotential $W=XY^2+Y^{k+1}+V_{\pm}$.
By the help of this confining duality we remove either the linear monopole  (or anti-monopole) superpotential
or both from the $U(\nn)$ model. In the first case we obtain a new duality, that reduces in the limiting case to the 
conjectured confining duality discussed above. In the second case we find the pure 3d duality obtained in \cite{Hwang:2018uyj} by reducing the 4d  Brodie duality with $U(\nn)$ gauge group.

Eventually we can also remove the KK monopole superpotential from the $Usp(2\nn)$ duality, finding another new duality
for pure 3d $Usp(2\nn)$ with two rank-two antisymmetric tensors.
Also in this last case we observe that the confining sector, obtained after the dual Higgsing, corresponds 
to the limiting case of the 3d $U(\nn)$ duality with a linear  monopole superpotential turned on.

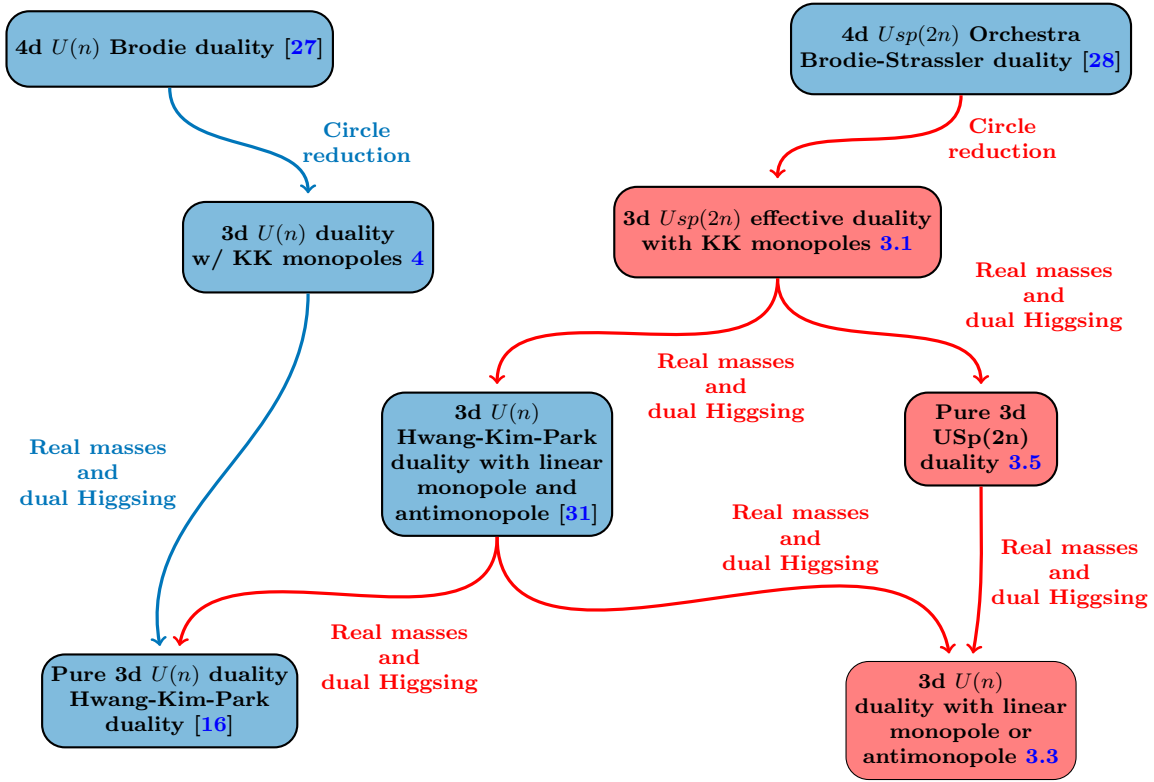
\begin{figure}
\begin{center}
\begin{tikzpicture}
\scriptsize\bfseries
	\node[draw, thick,
		fill=azzurro!50,
		rounded corners=3mm,
		minimum width=2cm, minimum height = 1cm] 
		(brodie) at (0,0) { 4d $U(\nn)$ Brodie duality \cite{Brodie:1996vx}};
	\node[draw, thick,
		fill=azzurro!50,
		rounded corners=3mm,
		minimum width=2cm, minimum height = 1.2cm,
		below right=1.5cm and -2cm of brodie] 
		(UKK)  {\makecell{3d $U(\nn)$ duality \\ w/ KK monopoles  \ref{sec:brodie43}}};
	\node[draw, thick,
		fill=azzurro!50,
		rounded corners=3mm,
		minimum width=2cm, minimum height = 1.2cm,
		below=7.5cm of brodie] 
		(U)  {\makecell{Pure 3d $U(\nn)$ duality \\ Hwang-Kim-Park \\ duality \cite{Hwang:2018uyj}}};
	\node[draw, thick,
		fill=azzurro!50,
		rounded corners=3mm,
		minimum width=2cm, minimum height = 1.2cm,
		right=6cm of brodie] 
		(brodie_strassler)  {\makecell{4d $Usp(2\nn)$ Orchestra \\ Brodie-Strassler duality \cite{Brodie:1996xm} }};
	\node[draw, thick,
		fill=rosso!50,
		rounded corners=3mm,
		minimum width=2cm, minimum height = 1.2cm,
		below left=1.2cm and -2cm of brodie_strassler] 
		(UspKK)  {\makecell{3d $Usp(2\nn)$ effective duality \\ with KK monopoles  \ref{subsec:uspKK}}};
	\node[draw, thick,
		fill=rosso!50,
		rounded corners=3mm,
		minimum width=2cm, minimum height = 1.2cm,
		below right=1.5cm and -0.5cm of UspKK] 
		(Usp)  {\makecell{Pure 3d \\ USp(2\nn) \\ duality \ref{subsec:noTUSp}}};
	\node[draw, thick,
		fill=azzurro!50,
		rounded corners=3mm,
		minimum width=2cm, minimum height = 1.2cm,
		below left=1.5cm and 0cm of UspKK] 
		(UT+T-)  {\makecell{3d $U(\nn)$ \\Hwang-Kim-Park \\ duality with linear \\ monopole and \\ antimonopole \cite{Hwang:2022jjs}}};
	\node[draw, 
		fill=rosso!50,
		rounded corners=3mm,
		minimum width=2cm, minimum height = 1.2cm,
		below =7.5cm of brodie_strassler] 
		(UT+)  {\makecell{3d $U(\nn)$ \\  duality with linear \\ monopole or \\ antimonopole  \ref{subsec:uTp}}};
	\draw[shorten >=0.1cm,->,very thick,azzurro] 
		(brodie) to[out=-90, in=90] node[right, xshift=0.7cm, azzurro] 
		{\makecell{Circle \\ reduction}} (UKK) ;
	\draw[shorten >=0.1cm,->,very thick,azzurro] 
		(UKK) to[out=-90, in=100] node[left, xshift=-0.5cm, azzurro] 
		{\makecell{Real masses \\ and \\ dual Higgsing}} (U) ;
	\draw[shorten >=0.1cm,->,very thick,rosso] 
		(brodie_strassler) to[out=-90, in=90] node[right, xshift=0.9cm, rosso] 
		{\makecell{Circle \\ reduction}} (UspKK) ;
	\draw[shorten >=0.1cm,->,very thick,rosso] 
		(UspKK) to[out=-90, in=90] node[right, yshift=-0.7cm, rosso] 
		{\makecell{Real masses \\ and \\ dual Higgsing}} (UT+T-) ;
	\draw[shorten >=0.1cm,->,very thick,rosso] 
		(UspKK) to[out=-90, in=90] node[right, xshift=1cm, yshift=0.5cm,  rosso] 
		{\makecell{Real masses \\ and \\ dual Higgsing}} (Usp) ;
	\draw[shorten >=0.1cm,->,very thick,rosso] 
		(UT+T-) to[out=-90, in=100] node[right, yshift=0.8cm, rosso] 
		{\makecell{Real masses \\ and \\ dual Higgsing}} (UT+) ;
	\draw[shorten >=0.1cm,->,very thick,rosso] 
		(UT+T-) to[out=-90, in=80] node[below, xshift=0.7cm, yshift=-0.2cm, rosso] 
		{\makecell{Real masses \\ and \\ dual Higgsing}} (U) ;
	\draw[shorten >=0.1cm,->,very thick,rosso] 
		(Usp) to[out=-90, in=80] node[right, xshift=0cm, yshift=-0cm, rosso] 
		{\makecell{Real masses \\ and \\ dual Higgsing}} (UT+) ;

\end{tikzpicture}
\caption{Survey of the dualities and the flows studied in this paper. The blue boxes and arrows represents 
dualities and flows that have already been proposed and studied in the literature. The red boxes and 
arrows represent the dualities and the flows proposed and analyzed here. }
\label{ease}
\end{center}
\end{figure}

For  ease of  reading we have summarized the various flows and dualities illustrated in this introduction in 
figure \ref{ease}.
We stress that this figure can be compared with the ones appearing in \cite{Amariti:2018wht} for the cases of $U(\nn)$ SQCD and $Usp(2\nn)$ SQCD with an adjoint. The search for such a homogenous picture for the web generated by the 
reduction of 4d Brodie and Brodie-Strassler duality was one of the main motivations of our work.


 A central role in our analysis is played by the study of the 4d superconformal index and by the 3d
 three sphere partition function \cite{Kapustin:2009kz,Jafferis:2010un,Hama:2010av,Hama:2011ea}. 
 Indeed we will show that each step taken on the field theory side can be reproduced by localization.
 Nevertheless, as already stressed in the case with a single adjoint, the 4d identities relating the superconformal indices of
 Brodie and Brodie-Strassler dualities are conjectural. 
 At large $n$ the 4d superconformal index of \cite{Kinney:2005ej} has been matched in \cite{Kutasov:2014wwa}, while a complete proof of such relation still lacks for finite $n$.

Anyway, starting from such conjectural identities we can obtain the 3d identities for the $U(\nn)$ and the $Usp(2\nn)$
cases with a KK monopole turned on, using the procedure of \cite{Aharony:2013dha}.
All the other steps, corresponding to real mass flow and Higgs flow on the field theory side, can be studied on the partition function by considering large limits for some of the parameters. If the divergent terms cancel in the identities we remain with
a finite result that can be interpreted as the new duality obtained after the flow.

In almost all of the flows analyzed here we have to provide an Higgs flow on the dual side. This reflects in a  large mass limit for the parameters associated to the gauge algebra, corresponding to the real scalar in the vector multiplet.
This is a delicate step, because it requires to commute limits and integral and some care is necessary. 
The new gauge sector created by such a dual Higgsing are then usually associated to confining gauge theories. Locally dualizing these sectors corresponds to computing the associated integrals. A very common feature of this computation is that the singlets of the confining sectors correspond to monopoles of the electric theory. In this way we obtain the final identities that correspond to the ones for the new 3d dualities.

In all of the derivation we have conjectured  an identity, relating $U(k-1)$ with two adjoints and a linear monopole 
superpotential and a set of (interacting) singlets. As a consistency check we have obtained this identity as a limiting case of
a more general duality and  we have recovered the integral identities for the partition functions for dualities already checked in terms of the superconformal index in the literature.

The paper is organized as follows. In section \ref{review} we give a short review of useful material necessary for our analysis.
In sub-section \ref{subsec:4d} we review the 4d dualities for $SU(\nn)$ SQCD with two adjoints and $Usp(2\nn)$ SQCD
with two rank-two antisymmetric tensors. In sub-section \ref{subsec:3d} we review the dualities discussed in \cite{Hwang:2018uyj} and \cite{Hwang:2022jjs} for 3d $U(\nn)$ SQCD with two adjoints. These two sub-sections exhaust the blue boxes in Figure \ref{ease}.
We conclude this review part with sub-section \ref{subsec:Z3d} where we report useful results from the analysis of 3d $\mathcal{N}=2$ gauge theories in terms of the three sphere partition function.
Section \ref{sec:uspred} is the core of the paper, where most of the new results are derived. 
We start by studying the circle reduction of $Usp(2\nn)$ SQCD with two rank-two antisymmetric tensors
in subsection \ref{subsec:uspKK} and providing the expected identity for the three sphere partition functions.
 Then in sub-section \ref{subsec:uTpTm} we flow from this effective duality of \cite{Hwang:2022jjs} with linear monopole and anti-monopole superpotential.
 In sub-section \ref{subsec:uTp} we conjecture a new confining duality and use this result to claim the existence of 
 another duality with a single linear monopole or anti-monopole superpotential.
 The conventional 3d duality of \cite{Hwang:2018uyj} is then obtained in section \ref{subsec:noTU} by a further real mass flow.
 We conclude the analysis in sub-section \ref{subsec:noTUSp} by removing the monopole superpotential from the 3d
 effective $Usp(2\nn)$ duality finding a new 3d duality for $Usp(2\nn)$ SQCD with two rank-two antisymmetric tensors
that generalizes the usual Aharony duality for $Usp(2\nn)$ SQCD.
In section \ref{sec:brodie43} we re-consider the circle compactification discussed in \cite{Hwang:2018uyj}, showing that it is possible to obtain  the various results without breaking  the gauge group in pure SQCD sectors. We corroborate the results  analyzing the various steps with the help of the three sphere partition function.
 In section \ref{sec:conc} we conclude by discussing many possible future directions. 
 
%
%
%
%
\section{Review}
\label{review}
%
%
%
%

%
%
%
%
\subsection{The 4d dualities}
\label{subsec:4d}
%
%
%
%
The original duality of \cite{Brodie:1996vx} relates
\begin{itemize}
\item 4d $SU(\nn)$ SQCD with $F$ flavors $Q$ and $\widetilde Q$ with two adjoints $X$ and $Y$ 
interacting through the superpotential
\begin{equation}
\label{Brodieel}
W = \Tr X Y^2 + \Tr X^{k+1}
\end{equation}
with $k$ odd.
\item 4d $SU(\tilde{n}=3kF-n)$ SQCD with $F$ dual flavors $q$ and $\widetilde q 
$ with two adjoints $x$ and $y$ 
interacting through the superpotential
\begin{equation}
\label{Brodiemag}
W = \Tr x y^2 + \Tr x^{k+1} + \sum_{j=0}^{k-1} \sum_{\ell=0}^{2} \Tr \mathcal{M}_{j,\ell} q x^{k-1-j} y^\ell \widetilde q
\end{equation}
where the singlets $\mathcal{M}_{j,\ell}$ correspond under the duality map to the dressed mesons  $Q X^j Y^\ell \tilde Q$ for $j=0,\dots,k-1$ and $\ell=0,1,2$ of the electric phase.
\end{itemize}

The non anomalous global symmetry is $SU(f)_L \times SU(f)_R \times U(1)_B \times U(1)_R$. The fields transform under the gauge and global symmetries as follows:
\begin{eqnarray}
\begin{array}{|c|c|c|c|c|c|c|}
\hline
\text{Field} & SU(\nn) &SU(\tilde n)& SU(f)_L & SU(f)_R & U(1)_B & U(1)_R \\
\hline
Q                 &n	        &1			&f	&	1	&	1	&	1-\frac{n}{f(k+1)}    \\
\hline \rule{0pt}{13pt} 
\tilde Q        &   \overline n        &1			&1	&	\overline f	&\!\!\!\!	-1	&		1-\frac{n}{f(k+1)}     \\
\hline
X                 &n^2-1	&1			&1	&	1	&	0	&	\frac{2}{k+1}    \\
\hline
Y                 &n^2-1	&1			&1	&	1	&	0	&	\frac{k}{k+1}    \\
\hline
q                 &. 1	&\overline{\tilde n}		&\overline f	&	1	&	 \frac{n}{\tilde n}&1-\frac{\tilde n}{f(k+1)}     \\
\hline
\tilde q         &  1   	&\tilde n 		&1		&	f	&	\!\!\!\!-\frac{n}{\tilde n}&	1-\frac{\tilde n}{f(k+1)}     \\
\hline
x                  &1		&\tilde n^2-1	&1		&1		&	0	&	\frac{2}{k+1}      \\
\hline
y                  &1		&\tilde n^2-1	&1		&1		&	0	&	\frac{k}{k+1}     \\
\hline
\mathcal{M}_{j \ell}      &  1        &1			&	f	&	\overline f	&	0	&	2-\frac{2}{k+1} +\frac{2(\ell-1)+k(j-1)}{k+1}   \\
\hline
\end{array}
\nonumber \\
\end{eqnarray}
Here we are actually interested in the $U(\nn)$ version of this duality, which is obtained by the standard procedure of gauging the baryonic symmetry, normalized opportunely, in both the electric and in the magnetic phase.

The other 4d dualities necessary for our analysis  was found in \cite{Brodie:1996xm} and it relates 
\begin{itemize}
\item
$Usp(2\nn)$ SQCD with $2 f$ fundamentals $Q$ and two rank-two antisymmetric 
tensors $A$ and $B$ interacting through a superpotential
\begin{equation}
\label{Worch}
W = Tr A^{k+1} + Tr A B^2 
\end{equation}
\item $USp(2\tilde{n}=2(3k f - n -4k-2))$ SQCD with $2f$ fundamentals $q$,
two rank-two antisymmetric \footnote{Here we refer to the reducible antisymmetric representation, made out of
an irreducible traceless tensor and a singlet. See \cite{Kapustin:2011vz} for a more complete discussion on this subtlety.} 
tensors $a$ and $b$ and the dressed mesons
$M_{rs}^{(j,\ell)} = Q_r A^j B^\ell Q_s$
symmetric (antisymmetric) for $j \ell$ odd  (even)
with $j=0,\dots,k-1$ and $\ell=0,1,2$.
The superpotential of the dual phase is
\begin{equation}
\label{Worchdual}
W = Tr  a^{k+1} + Tr a b^2+\sum_{j=0}^{k-1} \sum_{\ell=0}^{2} M_{j \ell} q  a^{k-j-1}  b^{2-\ell} q 
\end{equation}
where the singlets $\mathcal{M}_{j,\ell}$ correspond under the duality map to the dressed mesons  $Q X^j Y^\ell  Q$ for $j=0,\dots,k-1$ and $\ell=0,1,2$ of the electric phase.
Furthermore $k$ here is required to be odd, otherwise the global anomalies do not match.
\end{itemize}

The non anomalous global symmetry is $SU(2f) \times U(1)_B \times U(1)_R$. The fields transform under the gauge and global symmetries as follows:
\begin{eqnarray}
\begin{array}{|c|c|c|c|c|}
\hline
\text{Field} & Usp(2\nn) & USp(2 \tilde n) & SU(2f) & U(1)_R \\
\hline
Q &n&1&2f& 1-\frac{n+2k+1}{f(k+1)}\\
\hline
A &n(2n-1)&1&1&\frac{2}{k+1} \\ 
\hline
B &n(2n-1)&1&1&\frac{k}{k+1} \\
\hline
q&1&\tilde n&\overline{2f}& 1-\frac{n\tilde +2k+1}{f(k+1)}\\
\hline
a&1&\tilde n(2\tilde n-1)&1&\frac{2}{k+1} \\ 
\hline
b&1&\tilde n(2\tilde n-1)&1&\frac{k}{k+1} \\ 
\hline
\mathcal{M}_{j \,0} \, (j=0,\dots,k-1)&1&1& f(2f-1)& 2-  \frac{2(n+2k+1)}{f(k+1)}+\frac{2j}{k+1} \\
\hline
\mathcal{M}_{2j\, 1} \,(j=0,\dots,\frac{k-1}{2})&1&1&  f(2f-1)&  2-  \frac{2(n+2k+1)}{f(k+1)}+\frac{4j+k}{k+1}\\
\hline
\mathcal{M}_{2j+1\, 1} \,(j=0,\dots,\frac{k-3}{2})&1&1& f(2f+1) &2-  \frac{2(n+2k+1)}{f(k+1)}+\frac{4j+k+2}{k+1} \\
\hline
\mathcal{M}_{j \,2} \,(j=0,\dots,k-1)&1&1&  f(2f-1)& 2-  \frac{2(n+2k+1)}{f(k+1)}+\frac{2j+2k}{k+1} \\
\hline
\end{array} \nonumber \\
\end{eqnarray}

The 4d dualities discussed here can be translated in integral identities between the superconformal index of the electric and of the magnetic phase. 
These identities have been conjectured in \cite{Spiridonov:2009za} for the $SU(\nn)$ and the $Usp(2\nn)$ dualities.
The  identity for the $SU(\nn)$ duality corresponds to the equivalence between the integral
 \begin{eqnarray}
I_{SU( n)} &=&
\frac{(p,p)^{ n-1} (q,q)^{ n-1}}{  n!} \Gamma_e^{ n-1}(u) \Gamma_e^{ n-1} \big(u^\frac{k}{2} \big)  \\
&\times&
\int \prod_{\alpha=1}^{ n} \frac{d z_\alpha}{2\pi i z_\alpha} 
\prod_{a=1}^{f} \Gamma_e(s_a z_\alpha ) \Gamma_e(t_a^{-1} z_\alpha^{-1})
 \prod_{1\leq \alpha< \beta \leq  n} \frac{\Gamma_e(u (z_\alpha/z_\beta)^{\pm 1}) \Gamma_e(u ^\frac{k}{2} (z_\alpha/z_\beta)^{\pm 1})}{\Gamma_e( (z_\alpha/z_\beta)^{\pm 1})} \nonumber
\end{eqnarray}
with $\prod_{\alpha=1}^{n} z_\alpha=1$, $u=(pq)^{\frac{1}{k+1}}$ 
and  the integral
\begin{eqnarray}
I_{SU( \tilde n)} &=&
\frac{(p,p)^{\tilde n-1} (q,q)^{ \tilde n-1}}{  n!} \Gamma_e^{\tilde  n-1}(u) \Gamma_e^{\tilde  n-1} \big(u^\frac{k}{2} \big) 
 \prod_{j=0}^{k-1} \prod_{\ell=0}^{2} \prod_{a,b=1}^f \Gamma_e\big(u^\frac{2j+k \ell}{2} s_a/t_b\big)
  \\
&\times&
 \int \prod_{\alpha=1}^{ \tilde n} \frac{d z_\alpha}{2\pi i z_\alpha} 
\prod_{a=1}^{f} \Gamma_e(\tilde s_a z_\alpha^{-1} ) \Gamma_e(\tilde t_a^{-1} z_\alpha)
 \prod_{1\leq \alpha< \beta \leq  \tilde n} 
 \frac{\Gamma_e(u (z_\alpha/z_\beta)^{\pm 1}) \Gamma_e(u ^\frac{k}{2} (z_\alpha/z_\beta)^{\pm 1})}{\Gamma_e( (z_\alpha/z_\beta)^{\pm 1})} \nonumber
\end{eqnarray}
with $\prod_{\alpha=1}^{\tilde n} z_\alpha=1$,
 $\tilde s_\alpha = u^\frac {2-k}{2} \prod_{a=1}^{f} (s_a t_a)^\frac{3 k}{2n} s_a^{-1}$ 
 and
 $\tilde t_\alpha^{-1}= u^\frac {2-k}{2} \prod_{a=1}^{f} (s_a t_a)^{-\frac{3 k}{2n}} t_a$.
 
In this paper we skip most of the details of the index, referring the reader to \cite{Dolan:2008qi} for 
  conventions.
 For completeness we just report the  definitions  of the q-Pochammer symbols and of the elliptic Gamma functions 
  \begin{equation}
 (x;p) \equiv \prod_{j=0}^\infty (1-x p^j), \quad
 \Gamma_e(z) \equiv \prod_{j,k=0}^{\infty} \frac{1-p^{j+1} q^{k+1}/z}{1-z p^j q^k}
  \end{equation}
The arguments appearing in these functions are fugacities associated to the symmetries. For example $p$ and $q$ are associated to the isometries of $S^3$, $z_\alpha$ are associated to the gauge symmetry and the other fugacities refer
 to the matter fields. The fugacities are constrained by the anomaly cancellation as
\begin{equation}
\label{bcele4d}
u^n \prod_{a=1}^{f} s_a t_a^{-1} = (pq)^f
\end{equation}
This relation, commonly referred as balancing condition, is required for the validity of the integral identity.
Observe that here we will be interested in the $U(\nn)$ version of the duality.,
As explained in \cite{Aharony:2013dha} the  $U(\nn)$ identity can be obtained by gauging 
the baryonic symmetry, i.e. by integrating over the corresponding fugacity 
and by turning on the contribution of the FI term.

Similarly the $Usp(2\nn)$ duality corresponds to the equivalence between the integral 
\begin{eqnarray}
\label{USP4Iel}
I_{USp(2 n)} &=&
\frac{(p,p)^{ n} (q,q)^{ n}}{2^{ n}  n!} \Gamma_e^{ n}(u) \Gamma_e^{ n} \big(u^\frac{k}{2} \big) \nonumber \\
&\times&
\int \prod_{\alpha=1}^{ n} \frac{d z_\alpha}{2\pi i z_\alpha} \frac{\prod_{a=1}^{2f} \Gamma_e(s_a z_\alpha^{\pm 1} )}{\Gamma_e(z_\alpha^{\pm 2})} \prod_{1\leq \alpha< \beta \leq  n} \frac{\Gamma_e(u z_\alpha^{\pm 1} z_\beta^{\pm 1}) \Gamma_e(u ^\frac{k}{2} z_\alpha^{\pm 1} z_\beta^{\pm 1})}{\Gamma_e( z_\alpha^{\pm 1} z_\beta^{\pm 1})}
 \nonumber \\
\end{eqnarray}
with the balancing condition 
\begin{equation}
\label{balusp}
u^{n+2k+1} \prod_{a=1}^{2 f} s_a = (pq)^{f}
\end{equation}
and the integral 
\begin{eqnarray}
\label{USP4Imag}
I_{USp(2\tilde n)} &=&
\frac{(p,p)^{\tilde n} (q,q)^{\tilde n}}{2^{\tilde n} \tilde n!} \Gamma_e^{\tilde n}(u) \Gamma_e^{\tilde n} \big(u^\frac{k}{2} \big) \prod_{j=0}^{k-1} \prod_{\ell=0}^{2} \prod_{a<b} \Gamma_e\big(u^\frac{2j+k \ell}{2} s_a s_b\big) \prod_{q=0}^\frac{k-3}{2}\prod_{a=1}^{2f} \Gamma_e \big(u^{2q+1+\frac{k}{2}} s_a^2 \big)
\nonumber \\
&\times&
\int \prod_{\alpha=1}^{\tilde n} \frac{d z_\alpha}{2\pi i z_\alpha} \frac{\prod_{a=1}^{2f} \Gamma_e(u^{1-\frac{k}{2}}/s_a z_\alpha^{\pm 1} )}{\Gamma_e(z_\alpha^{\pm 2})} \prod_{1\leq \alpha< \beta \leq \tilde n} \frac{\Gamma_e(u z_\alpha^{\pm 1} z_\beta^{\pm 1}) \Gamma_e(u ^\frac{k}{2} z_\alpha^{\pm 1} z_\beta^{\pm 1})}{\Gamma_e( z_\alpha^{\pm 1} z_\beta^{\pm 1})} \nonumber \\
\end{eqnarray}
%
%
%
%
\subsection{The 3d dualities}
\label{subsec:3d}
%
%
%
%

The 3d duality obtained in  \cite{Hwang:2018uyj} from the dimensional reduction of Brodie duality relates
\begin{itemize}
\item 3d $U(\nn)$ SQCD with $f$ flavors $Q$ and $\widetilde Q$ with two adjoints $X$ and $Y$ 
interacting through the superpotential
\begin{equation}
\label{HKPel}
W = \Tr X Y^2 + \Tr X^{k+1}
\end{equation}
with $k$ odd.
\item 3d $U(\tilde n = 3k f - n )$ SQCD with $f$ dual flavors $q$ and $\widetilde q 
$ with two adjoints $x$ and $y$ 
interacting through the superpotential
\begin{eqnarray}
\label{Wdualfinal}
W &=& Tr x y^2+ Tr x^{k+1} +\sum_{j=0}^{k-1} \sum_{\ell=0}^{2} \Tr \mathcal{M}^{j,\ell} q x^{k-1-j}y^{2-\ell} \widetilde q
\nonumber \\
&+& \sum_{j=1}^{k-1} V_{j,0}^\pm {\widetilde V}_{k-j,0}^\pm
+\sum_{\ell=0}^{2} V_{0,\ell}^\pm {\widetilde V}_{0,2-\ell}^\pm
+ \sum_{q=0}^{\kappa=\frac{k-3}{2}}  W_{q}^\pm {\widetilde W}_{\kappa-q}^\pm
\end{eqnarray}
where the mesons correspond to the one discussed on the 4d side. There are also dressed monopoles operators
of the electric theory, 
carrying topological  charge $1$ and $2$,  denoted as 
$V_{j,\ell}^{\pm}$ and $W_{q}^{\pm}$ respectively,    acting as singlets in the dual phase, interacting through (\ref{Wdualfinal}) 
 with the dual dressed monopoles, carrying topological  charge $1$ and $2$, denoted as $\widetilde V_{j,\ell}^{\pm}$ and  $\widetilde W_{q}^{\pm}$ respectively.
These monopole operators are defined by radial quantization from states on $S^2$ carrying a non trivial magnetic flux background. The dressed monopole are then mapped to the following states
\begin{eqnarray}
\label{mondef}
V_{j \ell}^{\pm}  &\leftrightarrow& Tr X^j Y^\ell |\pm 1,0,\dots,0 \rangle \nonumber \\
W_q^{\pm}  &\leftrightarrow& Tr X^{2q} |\pm1,\pm1,0,\dots,0 \rangle
\end{eqnarray}
 and an analogous definition is given for the monopoles appearing in the dual phase.
As stressed in   \cite{Hwang:2018uyj}   the appearance of the $W_q^{\pm}$ and ${\widetilde W}_q^{\pm}$ monopoles is one of the most interesting novelties of models with two adjoints that does not have a counterpart in any supersymmetric duality worked out so far.
\end{itemize}

The  global symmetry is $SU(f)_L \times SU(f)_R \times U(1)_A \times U(1)_T \times U(1)_R$. 
The fields and the monopoles transform under the gauge and global symmetries as follows:
\begin{eqnarray}
\label{chU}
\begin{array}{|c|c|c|c|c|c|c|c|}
\hline
\text{Field} & U(\nn) &U(\tilde n)& SU(f)_L & SU(f)_R & U(1)_A & U(1)_T  & U(1)_R \\
\hline
Q                 &n	        &1			&f	&	1	&	1	&0	&r    \\
\hline
\tilde Q        &   \overline n        &1			&1	&	\overline f	& 1	&0	&	r    \\
\hline
X                 &n^2	&1			&1	&	1	&	0&0	&	\frac{2}{k+1}    \\
\hline
Y                 &n^2	&1			&1	&	1	&	0	&0	&	\frac{k}{k+1}    \\
\hline
V_{j\ell}^{\pm} &1 &1& 1&1&-f&\pm 1 & (1-r)f+\frac{2j + k \ell -(n-1)}{k+1} \\
\hline
W_{q}^{\pm} &1 &1& 1&1&-2f&\pm 2 & 2(1-r)f+\frac{2 + 4 q  -2(n-1)}{k+1} \\
\hline
q                 & 1	&\overline{\tilde n}		&\overline f	&	1&	\!\!\!\!-1	& 0&\frac{2-k}{k+1} -r    \\
\hline
\tilde q         &  1   	&\tilde n 		&1		&	f	&	\!\!\!\!-1&	0&	\frac{2-k}{k+1} -r     \\
\hline
x                  &1		&\tilde n^2	&1		&1		&	0	&	0&	\frac{2}{k+1}      \\
\hline
y                  &1		&\tilde n^2	&1		&1		&	0	&0	&	\frac{k}{k+1}     \\
\hline
\mathcal{M}_{j \ell}      &  1        &1			&	f	&	\overline f	&	2	& 0	&2r+\frac{2j+k \ell}{k+1}  \\
\hline \rule{0pt}{13pt} 
{\widetilde  V}_{j\ell}^{\pm} &1 &1& 1&1&f&\mp 1 & (r-1)f+\frac{2j + k \ell +(n+1)}{k+1} \\
\hline \rule{0pt}{13pt} 
{\widetilde W}_{q}^{\pm} &1 &1& 1&1&2f&\mp 2 & 2(r-1)f+\frac{2 + 4 q  +2(n+1)}{k+1} \\
\hline
\end{array} \nonumber \\
\end{eqnarray}

Another useful duality was obtained in the more recent paper \cite{Hwang:2022jjs}.
It relates 
\begin{itemize}
\item 3d $U(\nn)$ SQCD with $F$ flavors $Q$ and $\widetilde Q$ with two adjoints $X$ and $Y$ 
interacting through the superpotential
\begin{equation}
\label{HK2Pel}
W = \Tr X Y^2 + \Tr X^{k+1} + V_{0,0}^+ +  V_{0,0}^-
\end{equation}
with $k$ odd.
\item 3d $U(\tilde n = 3k f-n-4k-2)$ SQCD with $F$ dual flavors $q$ and $\widetilde q 
$ with two adjoints $x$ and $y$ 
interacting through the superpotential
\begin{eqnarray} 
\label{W2dualfinal}
W &=& Tr x y^2+ Tr x^{k+1} +\sum_{j=0}^{k-1} \sum_{\ell=0}^{2} \Tr \mathcal{M}^{j,\ell} q x^{k-1-j}y^{2-\ell} \widetilde q
+ \widetilde V_{0,0}^+ + \widetilde V_{0,0}^-
\end{eqnarray}
where the mesons are defined as above
\end{itemize}
In this case the monopole superpotentials (\ref{HK2Pel}) and (\ref{W2dualfinal})  break the axial and  the topological symmetry. Furthermore they fix the $R$-charge $r$ in (\ref{chU}) to $r=1-\frac{n+2k+1}{k+1}$, that can be computed 
from $R[V_{0,0}] = R[{\widetilde V} _{0,0}]=2$.

Another duality studied in \cite{Hwang:2022jjs} , involves $U(\nn)$ with the linear  monopole and anti-monopole superpotential for $W_{0,0}^{\pm}$ and ${\widetilde W}_{0,0}^{\pm}$
\begin{equation}
\label{HKrPel}
W = \Tr X Y^2 + \Tr X^{k+1} + W_{0,0}^+ +  W_{0,0}^-
\end{equation}
and a dual $U(3kF-\nn-2k-2)$ gauge theory 
 with superpotential
\begin{eqnarray}
\label{W3dualfinal}
W &=& Tr x y^2+ Tr x^{k+1} +\sum_{j=0}^{k-1} \sum_{\ell=0}^{2} \Tr \mathcal{M}^{j,\ell} q x^{k-1-j}y^{2-\ell} \widetilde q
+ \widetilde W_{0,0}^+ + \widetilde W_{0,0}^-
\end{eqnarray}
Here we will not consider this case, even it may be interesting to derive this duality flowing from some of the
other cases studied here.

%
%
%
%
\subsection{3d partition function}
\label{subsec:Z3d}
%
%
%
%
 
 In this section we report some useful facts on the analysis of the three sphere partition function that 
 play a central role in our analysis. 
The partition function computed on the squashed three sphere is given by a matrix integral over a vector  $\vec \sigma$ representing the scalar field in the Cartan of the gauge group.
The functions that appear in the integrand are the one loop determinants of the chiral and vector multiplets and they can be formulated in terms of hyperbolic Gamma functions. These are defined as
\begin{equation}
  \label{hvbd}
  \Gamma_h(x;\omega_1,\omega_2) \equiv
  \Gamma_h(x)\equiv 
  e^{
    \frac{\pi i}{2 \omega_1 \omega_2}
    \left((x-\omega)^2 - \frac{\omega_1^2+\omega_2^2}{12}\right)}
  \prod_{j=0}^{\infty} 
  \frac
  {1-e^{\frac{2 \pi i}{\omega_1}(\omega_2-x)} e^{\frac{2 \pi i \omega_2 j}{\omega_1}}}
  {1-e^{-\frac{2 \pi i}{\omega_2} x} e^{-\frac{2 \pi i \omega_1 j}{\omega_2}}}.
\end{equation}
where $\omega_1=i b$ and $\omega_2=i/b$ are related to the squashing parameter of the three sphere, defined by the equation $b^2(x_1^2+x_2^2) + b^{-2} (x_3^2+x_4^2)=1$. The parameter $\omega$ is defined as $2\omega= \omega_1+\omega_2$.
The other argument corresponds to the sum of the weights of the representation of each field with respect to the scalar in the gauge multiplet $\vec \sigma$ and the scalar obtained by weakly gauging the flavor symmetries (corresponding to the real masses in the field theory language).
There are also classical contributions to the integral, arising from CS and FI terms. 
These results are standard in the literature and for this reason we skip most of the definitions. Here we use the 
conventions of  \cite{Benini:2011mf} and we refer to that paper for further details.

Focusing on the models of interest here,
the partition function for $Usp(2\nn)$ SQCD with $2f$ fundamentals and two antisymmetric $A$ and $B$ is given by
\begin{eqnarray}
Z_{Usp(2\nn)}^{2 f} (\vec \mu;\tau_A,\tau_B)
&=&
\frac{\Gamma_h(\tau_A)^n \Gamma_h(\tau_B)^n}{{\sqrt{- \omega_1 \omega_2}}^n 2^n n!}
\int \prod_{i=1}^{n} d\sigma_i
\frac{\prod_{a=1}^{2f}
\Gamma_h(\pm \sigma_i + \mu_a)}
{\Gamma_h (\pm 2 \sigma_i)}
\nonumber \\
&\times&
\prod_{i<j} \frac{\Gamma_h(\pm \sigma_i + \pm \sigma_j + \tau_A)\Gamma_h(\pm \sigma_i + \pm \sigma_j + \tau_B)}
{\Gamma_h(\pm \sigma_i \pm \sigma_j)}
\end{eqnarray}
The superpotential $W=A^{k+1}+A B^2$  fixes as $\tau_A = \frac{2 }{k+1}\omega$ and $\tau_A = \frac{k }{k+1} \omega$.
Furthermore in presence of a KK monopole superpotential we must impose the further constraints
\begin{equation}
\label{bc1}
\sum_{a=1}^{2f} \mu_a = 2(\omega(f-2)-(n-1) (\tau_A+\tau_B-\omega)) 
\end{equation}
On the other hand the partition function for $U(\nn)$ SQCD with $f$ pairs of fundamentals and anti-fundamentals and two adjoints $X$ and $Y$ is given by
\begin{eqnarray}
Z_{U(\nn)}^{f} (\vec \mu;\vec v;\tau_X,\tau_Y;\xi)
&=&
\frac{\Gamma_h(\tau_X)^n \Gamma_h(\tau_Y)^n}{{\sqrt{- \omega_1 \omega_2}}^n n!}
\int \prod_{i=1}^{n} d\sigma_i e^{-2i \pi \xi \sigma_i}
\prod_{a=1}^{f}
\Gamma_h( \sigma_i + \mu_a)\Gamma_h(- \sigma_i + \nu_a)
\nonumber \\
&\times&
\prod_{i<j} \frac{\Gamma_h(\pm(\sigma_i -\sigma_j) + \tau_X)\Gamma_h(\pm (\sigma_i - \sigma_j) + \tau_Y)}
{\Gamma_h(\pm( \sigma_i - \sigma_j))}
\end{eqnarray}
The parameter $\xi$ is the FI, corresponding to the mass parameter for the topological $U(1)_T$ symmetry.
The superpotential $W=X^{k+1}+ XY^2$  fixes as $\tau_X = \frac{2 }{k+1}\omega$ and $\tau_Y = \frac{k }{k+1} \omega$.
In this case we can impose different constraints on the parameters, depending on the presence of a KK monopole superpotential or a linear monopole superpotential.
The KK monopole imposes 
\begin{equation}
\label{bcs1}
\sum _{a=1}^f \left(\mu _a+\nu _a\right)=2 f \omega -n \tau_X
\end{equation}
The linear  monopole and anti-monopole superpotential imposes $\xi=0$ and
\begin{equation}
\sum _{a=1}^f (\mu _a+\nu _a)= 2(\omega(f-2)-(n-1) (\tau_X+\tau_Y-\omega)) 
\end{equation}
while the linear  monopole (or anti-monopole) superpotential imposes the constraint
\begin{equation}
\label{bcmon1}
\sum _{a=1}^f (\mu _a+\nu _a)=+( \text{or } -) 2\xi+2 ((f-2) \omega - (n-1) \left(\tau _X+\tau _Y-\omega \right))
\end{equation}

We conclude this section by referring to two useful formulas that play a crucial role in our analysis.
The fist one is the inversion formula for the hyperbolic Gamma functions $\Gamma_h(2\omega-x) \Gamma_h(x)=1$.
This formula corresponds in  field theory to integrate  fields appearing in the superpotential through a holomorphic mass term.
The second formula is
\begin{equation}
\label{largerealmass}
\lim_{|x| \rightarrow \infty} \Gamma_h(x) = e^{-\frac{i \pi}{2} \text{sign}(x) (x-\omega)^2}
\end{equation}
and it corresponds to integrate out fields with a large real mass term. The CS contact terms generated by 
this operation follow from the gaussian factor in \eqref{largerealmass}. 
%
%
%
%
\section{4d/3d reduction of $USp(2\nn)$ SQCD with two rank-two anti-symmetric tensors}
\label{sec:uspred}
%
%
%
%
In this section we derive the 3d dualities starting from the circle compactification of the one for
$Usp(2\nn)$ SQCD with two rank-two antisymmetric tensors.
Before starting the analysis we have decided to simplify the reading by summarizing the final dualities  in (\ref{allinone}).
We have specified the rank of the electric and of the magnetic gauge group, the electric and the magnetic superpotential and the paragraph where we  have discussed the derivation of each duality.

\begin{eqnarray}
\label{allinone}
\begin{array}{|c|c|c|c|c|}
\hline
\text{G}_{\text{electric}}  &    \text{G}_{\text{magnetic}}  &   W_{\text{electric}}  &    W_{\text{magnetic}}  &  \text{Paragraph} \\
\hline
 Usp(2\nn)&  USp(2(3kf-n-4k-2))  & (\ref{Worcheta})&(\ref{Worchdualeta}) &  \ref{subsec:uspKK} \\
 \hline
 U(\nn)&  U(3kf-n-4k-2)  & (\ref{HK2Pel})&(\ref{W2dualfinal}) & \ref{subsec:uTpTm}\\
 \hline
 U(\nn) & U(3kf-n-2k-1) & (\ref{eleT})&(\ref{Wsingle}) & \ref{subsec:uTp}\\
 \hline
 U(\nn) & U(3kf-n)& (\ref{HKPel})&(\ref{Wdualfinal}) & \ref{subsec:noTU}\\
 \hline
 Usp(2\nn) & USp(2(3kf-n-2k-1)) & (\ref{Worch})&(\ref{finalUSpAha}) & \ref{subsec:noTUSp}\\
 \hline
 U(\nn) & U(3kf-n) &  (\ref{Brodieel})+(\ref{etael})&(\ref{Brodiemag})+(\ref{etamag}) & \ref{sec:brodie43} \\
    \hline
\end{array}
\nonumber \\
\end{eqnarray}

%
%
%
%
\subsection{$USp(2\nn)$ with KK monopole superpotential}
\label{subsec:uspKK}
%
%
%
%
The circle reduction of the 4d Brodie-Strassler duality for $Usp(2\nn)$ with $2f$ fundamentals and two adjoints
gives rise to the following 3d effective duality
\begin{itemize}
\item
3d $Usp(2\nn)$ SQCD with $2 f$ fundamentals $Q$ and two rank-two antisymmetric 
tensors $A$ and $B$
\begin{equation}
\label{Worcheta}
W = Tr A^{k+1} + Tr A B^2 + \eta Y_{USp}
\end{equation}
\item 3d $USp(2(3k f - n -4k-2))$ SQCD with $2f$ fundamentals $q$,
two rank-two antisymmetric 
tensors $a$ and $b$ and the dressed mesons
$M_{rs}^{(j,\ell)} = Q_r A^j B^\ell Q_s$
symmetric (antisimmetric) for $j \ell$ odd  (even)
with $j=0,\dots,k-1$ and $\ell=0,1,2$.
The superpotential of the dual phase is
\begin{equation}
\label{Worchdualeta}
W = Tr  a^{k+1} + Tr a b^2+\sum_{j=0}^{k-1} \sum_{\ell=0}^{2} M_{j \ell} q  a^{k-j-1}  b^{2-\ell} q 
+{\tilde \eta y}_{USp}
\end{equation}
with $k$ odd.
\end{itemize}
The  monopole operators $Y_{USp}$ and ${y}_{USp}$ can be generated because the rank-two antisymmetric tensors 
do not carry new zero modes in the KK monopole background \cite{Golkar:2009aq,Amariti:2018wht}.

The identity relating the superconformal indices of the 4d theory has been conjectured in \cite{Spiridonov:2009za}.
Reducing the identity between the indices to an identity between the  partition functions
can be done along the lines of the prescription given in \cite{Aharony:2013dha}. 
Such prescription requires a  re-definition of the fugacities $p$, $q$ associated to the isometries of $S^3$, of the fugacity $u_a$ associated to the global symmetry and of the fugacity $z_i$ associated to the gauge symmetry:
\begin{equation}
p = e^{2 \pi i r \omega_1},\quad q=e^{2 \pi i r \omega_2}, \quad z = e^{2 \pi i r \sigma},\quad u =  e^{2 \pi i r m}
\end{equation}
The 3d limit is taken as $r \rightarrow 0$, and it can be shown that the elliptic Gamma function become hyperbolic Gamma functions 
\begin{equation}
\lim_{r \rightarrow 0} \Gamma_e(e^{2 \pi i r z}) = e^{-\frac{i \pi}{6 \omega_1 \omega_2 r}(z-\omega)} \Gamma_h(z)
\end{equation}
There is a divergent contribution associated to the gravitational anomalies, that cancels if one reduces an integral identity between 4d dual models. 
By applying this reduction to the identity between (\ref{USP4Iel})  and (\ref{USP4Imag}) we arrive at
\begin{eqnarray}
\label{firstidusp}
Z_{Usp(2\nn)}^{2 f} (\vec \mu;\tau_A,\tau_B)
&=&
Z_{USp(2(3kf-n-4k-2))}^{2 f} (\tau_A-\tau_B-\vec \mu;\tau_A,\tau_B)
\nonumber \\
&\times&
\prod_{j=0}^{k-1}\prod_{\ell=0}^2 \prod_{1\leq a<b\leq f}  \Gamma_h(j \tau_A+\ell \tau_B+\mu_a+\mu_b)
\nonumber \\
&\times&
\prod_{q=0}^{k-1}\prod_{a=1}^f \Gamma_h((2q+1) \tau_A+\tau_B+2\mu_a) 
\end{eqnarray}
with the constraint (\ref{bc1}) among the mass parameters.
This constraint descends from the 4d constraint (\ref{balusp}).
While this constraint corresponds to anomaly cancellations  in 4d, in 3d it
reflects the presence of the KK monopole superpotential, both in the electric and in the magnetic phase, preventing 
the generation of an axial symmetry. 
This is the crucial aspects underlining the reduction of 4d dualities to 3d, the presence of the KK monopoles
generates new effective dualities on the circle, having the same global symmetries of the 4d ones.
This guarantees that the 4d  duality is preserved in the effective 3d description.

 It is then possible to flow to other 3d dualities by consistently removing the KK monopole superpotential in both phases. 
This is in general non trivial because the real mass flow necessary to remove the monopole superpotential 
in the electric phase can require also a dual Higgsing in the magnetic one. On the partition function this translates 
into the cancellation of the divergent pre-factors emerging when integrating out the massive field in both sides of the identities.
In the following we  study some of these flows starting from the duality with the KK monopole superpotential 
and from the identity (\ref{firstidusp}).
In this way we  obtain  the other 3d dualities summarized in figure (\ref{ease}) (except  for the $U(\nn)$ case with a
KK monopole turned on, that will be discussed separately in Section \ref{sec:brodie43}).

%
%
%
%
\subsection{$U(\nn)$ with linear  monopole and anti-monopole superpotential}
\label{subsec:uTpTm}
%
%
%
%
The first flow under investigation gives rise to the duality of \cite{Hwang:2022jjs} with linear 
monopole and anti-monopole superpotential reviewed in sub-section \ref{subsec:3d}.

The flow consists of assigning a large positive real mass to $f$ of the  fundamentals and 
a large negative and opposite real mass to the remaining $f$ fundamentals.
This real mass flow is then combined with a Higgs flow, corresponding to a shift of the real
scalars in the gauge group, equal (or equivalently opposite)  to the one assigned to $f$ fundamentals.

The electric theory becomes $U(\nn)$ with two adjoint and $f$ pairs of fundamentals and anti-fundamentals,
$Q$ and $\tilde Q$.
On the dual side the real scalars in the gauge group are shifted accordingly, giving rise to 
an $U(3k f-n-4k-2)$ gauge theory with $f$ pairs of fundamentals $q$ and $\tilde q$ and anti-fundamentals and two adjoints.
In this case only the antisymmetric contributions from the original mesons remains in the low energy spectrum and they correspond to the dressed mesons $M_{j,\ell}^{r,s} = Q^r X^j Y^\ell \tilde \tilde Q^s$.
The flow does not generate any axial and topological symmetry, reflecting the presence of the linear monopole and anti-monopole superpotential of \cite{Hwang:2022jjs}.

This duality can be obtained  studying the flow on the partition function as well. In this case we assign the masses
as
\begin{equation}
\left\{
\begin{array}{cccl}
\mu_a &=& m_a+s& \quad a=1,\dots,f
\\
\mu_a &=& n_a-s& \quad a=f+1,\dots,2f
\end{array}
\right.,
\end{equation}
and we shift the integration variables $\sigma_i \rightarrow \sigma_i+s$ in both the electric and the magnetic phase. 

By computing the large $s$ limit and by canceling the common  divergent pre-factors 
between the LHS and the RHS of the identity (\ref{firstidusp}) we obtain a new identity 
\begin{eqnarray}
\label{firstU}
Z_{U(\nn)}^{f} (\vec m; \vec n,;\tau_A,\tau_B;0)
&=&
Z_{U(3kf-n-4k-2)}^{f} (\tau_A-\tau_B-\vec n; \tau_A-\tau_B-\vec n ;\tau_A,\tau_B;0)
\nonumber \\
&\times&
\prod_{j=0}^{k-1}\prod_{\ell=0}^2 \prod_{a,b=1}^{f}  \Gamma_h(j \tau_A+\ell \tau_B+m_a+n_b)
\end{eqnarray}
with the constraint 
\begin{equation}
\label{bc2}
\sum_{a=1}^{f} (m_a+n_b) = 2(\omega(f-2)-(n-1) (\tau_A+\tau_B-\omega)) 
\end{equation}
This is the expected identity for the duality with linear  monopole and anti-monopole superpotential. 
The constraints imposed by the superpotential correspond indeed to the absence of an FI in 
(\ref{firstU}) and to the constraint (\ref{bc2}), as discussed in sub-section \ref{subsec:Z3d}.

%
%
%
%
\subsection{$U(\nn)$ with a single linear monopole superpotential term}
 \label{subsec:uTp}
%
%
%
%

The next step consist of integrating out one massive flavor from the last duality and flow to a new duality 
with a single one monopole (or anti-monopole) superpotential.
This flow requires some care in the magnetic sector, because it requires a dual Higgsing in order to match the 
discrete anomalies. 

If we consider $f+1$ flavors on the electric side and assign large opposite real masses to the last pair 
 on the dual side.
 The electric model becomes
 \begin{equation}
 \label{eleT}
 W = Tr X Y^2+ Tr X^{k+1} + V_{0,0}^{+}
 \end{equation}
 The linear  monopole (anti-monopole) superpotential $W_{mon} = V_{0,0}^+$
survives the real mass flow.
A completely equivalent choice is given by a linear superpotential for the anti-monopole, \emph{i.e.} 
$W_{mon}= V_{0,0}^-$. 
Keeping $V_{0,0}^+$ or $V_{0,0}^-$ in the superpotential is determined 
by the sign of the large real mass assigned to the fundamental quarks.

In the magnetic theory, after the dual Higgsing, we have a $U(3k f-n-2k-1) \times  U(k-1)$ gauge group.
 There are $f$ flavors $q$ and $\tilde q$ in the first gauge sector and $1$ pair $q_\rho$ and $\tilde q_\rho$
 in the magnetic one.
 There are also two pairs of adjoints,  $x$ and $y$ in the $U(3k f-n-2k-1) $ sector 
 and $ x_\rho$ and $ y_\rho$ in the $U(k-1)$ sector.
The superpotential is
\begin{eqnarray}
\label{Wsingle0}
W &=& Tr x y^2+ Tr x^{k+1} + Tr x_\rho y_\rho^2+ Tr x_\rho^{k+1} +\widetilde V_{0,0}^+ +
\widetilde{\mathbf{V}}_{0,0}^+
\nonumber \\
&+&\sum_{j=0}^{k-1} \sum_{\ell=0}^{2} \Tr (\mathcal{M}^{j,\ell} q x^{k-1-j}y^{2-\ell} \widetilde q +\mathcal{M}_\rho^{j,\ell} q_{\rho} x_\rho^{k-1-j} y_\rho^{2-\ell} \widetilde q_{\rho} )+
\nonumber \\
&+&   \sum_{j=1}^{k-1} \widetilde{V}_{j,0}^- \widetilde {\mathbf{V}}^-_{k-j,0}
+
\sum_{\ell=0}^{2} \widetilde V_{0,\ell}^- \widetilde {\mathbf{V}}^-_{0,2-\ell}
+
\sum_{q=0}^{\kappa=\frac{k-3}{2}}
{\widetilde W_{q}}^- \widetilde {\mathbf{W}}^-_{\kappa-q}
\nonumber \\
\end{eqnarray}
where the singlets $\mathcal{M}$ and $\mathcal{M}_\rho$  are the light field  
that survive from the original mesons in the real mass flow.
Furthermore the bare monopoles and anti-monopoles of the $U(k-1)$ sector have been denoted as $\widetilde{\mathbf{V}}^\pm_{0,0}$ and $\widetilde{\mathbf{W}}^\pm_{0,0}$

Consistently with the assignation of the masses in the electric theory 
the linear  monopole (anti-monopole) superpotential $W_{mon}=\widetilde{V}_{0,0}^+ +\widetilde {\mathbf{V}}_{0,0}^+$
(or $W_{mon}=\widetilde{V}_{0,0}^-+\widetilde {\mathbf{V}}_{0,0}^-$) remains in the superpotential.
On the other hand the last line in (\ref{Wsingle}) represents the AHW-like superpotential 
generated by the dual Higgsing between the monopole operators of the two gauge sectors.

We can consider this relation between the electric and the magnetic theory as a new duality for $U(\nn)$
SQCD with two adjoint an a single  monopole turned on.
Nevertheless here we are interested in a more conventional formulation of the dual model.
This is achieved by conjecturing that the $U(k-1)$ sector with two adjoints $x_\rho$ and $y_\rho$
interacting through $W=Tr x_\rho^{k+1} + Tr x_\rho y_\rho^2$ and
 one fundamental flavor identified with $q_\rho$ and $\tilde q_\rho$  is confining
 \footnote{Observe that in 4d the confining limiting case of Brodie duality has been studied in  \cite{Klein:1998uc}.
 In 3d the number of confining limiting cases is expected to be larger, because of the possible 
 presence of monopole superpotentials.} 
in presence of a linear monopole superpotential, either $W_{mon}=\widetilde {\mathbf{V}}_{0,0}^+$
(or $W_{mon}=\widetilde {\mathbf{V}}_{0,0}^-$).
In absence of further interactions the confining theory corresponds to a set of singlets identified with the 
dressed mesons $q_{\rho} x_\rho^{k-1-j} y_\rho^{2-\ell} \widetilde q_{\rho}$ 
and with the  monopoles $\widetilde {\mathbf{V}}_{j,\ell}^{-}$ (or $\widetilde {\mathbf{V}}_{j,\ell}^+$) with $j \ell = 0$ and $\widetilde{\mathbf{W}}_{q}^{-}$ (or  $\widetilde{\mathbf{W}}_{q}^{+}$) with $q-=0,\dots,\frac{k-3}{2}$.

This conjecture allows us to dualize the $U(k-1)$ sector and the dual model becomes $U(3kf-n-2k-1)$ with superpotential
  \begin{eqnarray}
\label{Wsingle}
W &=& Tr x y^2+ Tr x^{k+1}  +\widetilde V_{0,0}^+ +
\sum_{j=0}^{k-1} \sum_{\ell=0}^{2} \Tr (\mathcal{M}^{j,\ell} q x^{k-1-j}y^{2-\ell} \widetilde q)+
\nonumber \\
&+&   \sum_{j=1}^{k-1} \widetilde{V}_{j,0}^- V_{k-j,0}^-
+
\sum_{\ell=0}^{2} \widetilde V_{0,\ell}^-  V_{0,2-\ell}^-
+
\sum_{q=0}^{\kappa=\frac{k-3}{2}}
\widetilde W_{q}^-  W_{\kappa-q}^-
\nonumber \\
\end{eqnarray}
where in the last line we have identified the monopoles of the confining sector with some of the monopoles of the electric theory  namely $V_{j,0}^+$, $V_{0,\ell}^+$ and $W_{q}^+$.

We can translate the conjecture on the partition function side, conjecturing the following identity 
\begin{eqnarray}
\label{conj}
&&
Z_{U(k-1)}^{1}(m,m;\tau_A,\tau_B; 2(\tau _A-\tau _B- \omega-m))
=
\nonumber \\
 \times &&
\prod_{j=0}^{k-1}\prod_{\ell=0}^{2} 
\Gamma_h(2m+ j\tau_A +\ell \tau_B )
 \prod_{q=0}^{ \frac{k-3}{2}} \Gamma_h((2q+1) \tau_A - 4 (m-\tau_A+\tau_B))
\nonumber \\
 \times &&
 \prod_{j=0}^{k-1}\Gamma_h (j \tau_A - 2 (m-\tau_A+\tau_B)))
\prod_{\ell=1}^{2} \Gamma_h (\ell \tau_B - 2 (m-\tau_A+\tau_B)))
\end{eqnarray}
where the FI is fixed by the balancing condition (\ref{bcmon1}), corresponding to the presence of the linear monopole 
superpotential.

The flow is triggered by considering the identity (\ref{firstU}) with $f+1$ fundamental flavors and assigning 
a large mass parameter to  the last pair as
\begin{equation}
\left\{
\begin{array}{ccc}
m_{f+1} &=& \mathbf{m} +s
\\
n_{f+1} &=& \mathbf{m} -s
\end{array}
\right.,
\end{equation}
On the dual side the gauge group $U(3kf-n-k-2)$ is Higgsed to 
$U(3kf-n-2k-1) \times U(k-1)$ by shifting $(k-1)$ integration variables as $\sigma_i\rightarrow \sigma_i-s$.
Computing the large $s$ limit on the identity (\ref{firstU}) and eliminating the divergent part, that coincides on the electric and on magnetic side we end up with the relation
\begin{eqnarray}
\label{intermedT}
Z_{U(\nn)}^f(\vec m, \vec n;\tau_A;\tau_B;-2(\mathbf{m}-\omega))
&=&
Z_{U(3kf-n-2k-1)}^f(\vec n,\vec m ;\tau_A;\tau_B;2(\mathbf{m}-\tau_B))
\nonumber \\
&\times&
Z_{U(k-1)}^1(\tilde{\mathbf{m}},\tilde{\mathbf{m}};\tau_A,\tau_B;-2(\mathbf{m}-\omega))
\nonumber \\
&\times&
\prod_{j=0}^{k-1}\prod_{\ell=0}^2 
\prod_{a,b=1}^{f}  \Gamma_h(j \tau_A+\ell \tau_B+m_a+n_b) 
\nonumber \\
&\times&
\prod_{j=0}^{k-1}\prod_{\ell=0}^2 \prod_{a,b=1}^{f}  \Gamma_h(j \tau_A+\ell \tau_B+2m)
e^{ \frac{i \pi (k-1) }{2} \sum _{a=1}^{f} (m _a^2- n_a^2)}
\nonumber \\
\end{eqnarray}
 with $\vec {\tilde n} = \tau_A-\tau_B- \vec n$, 
 $\vec {\tilde m} = \tau_A-\tau_B- \vec m$ and
 $\tilde{\mathbf{m}}=\tau_A-\tau_B- \mathbf{m}$.
 There is also a constraint between the masses corresponding to
 \begin{equation}
 \sum _{a=1}^f (m_a+n_a)=2 ((f-1) \omega -(n-1) (\tau _A+\tau _B-\omega ))-2 \mathbf{m}
 \end{equation}
 that signals the presence of the monopole superpotential both in the $U(\nn)$ and
 in the $U(3kf-n-2k-1)$ sector. The monopole superpotential in the $U(k-1)$ sector 
 corresponds to the constraint on the FI in the second line of (\ref{intermedT}). 
 The integral corresponding to the $U(k-1)$ sector can then be computed using the conjectured 
 relation (\ref{conj}). After eliminating the massive fields with the inversion formula for the hyperbolic gamma functions
 we obtain the relation
 \begin{eqnarray}
\label{finalT}
Z_{U(\nn)}^f(\vec m, \vec n;\tau_A\!&\!;\!&\!\tau_B;2(\omega- \mathbf{m}))
=
Z_{U(3kf-n-2k-1)}^f(\vec n,\vec m ;\tau_A;\tau_B;2(\mathbf{m}-\tau_B))
\nonumber \\
&\times&
\prod_{j=0}^{k-1}\prod_{\ell=0}^2 
\prod_{a,b=1}^{f}  \Gamma_h(j \tau_A+\ell \tau_B+m_a+n_b) 
e^{ \frac{i \pi (k-1) }{2} \sum _{a=1}^{f} (m _a^2- n_a^2)}
\nonumber \\
&\times& \!\!\!\!\!\!\!\! \!\!
\prod_{\tiny{
\begin{array}{c}
j=0,\dots,k-1\\
\ell=0,\dots,2\\
j \ell=0
\end{array}
}}
\!\!\!\!\!\!\!\! \Gamma_h( (f-1) \omega-\frac{n-1}{2} \tau _A -\sum _{a=1}^f (m _a+n _a)+j \tau_A+\ell \tau_B)
\nonumber \\
&\times&
\prod_{q=0}^{\frac{k-3}{2}}
\Gamma_h(2 (f-1) \omega-(n-1) \tau _A -2\sum _{a=1}^f (m _a+n _a)+(2q+1) \tau_A)
\nonumber \\
\end{eqnarray}
This relation reproduces, on the partition function, the duality that we have claimed on the field theory side.
Furthermore the identity reduces to the one conjectured in formula (\ref{conj}) for $n=k-1$, $f=1$ and $m_1=n_1$. 
%
%
%
%
\subsection{$U(\nn)$ without  monopole superpotential}
 \label{subsec:noTU}
%
%
%
%
Another crucial sanity check of our construction consists of flowing to the pure 3d duality of \cite{Hwang:2018uyj}.
This can be achieved by starting from the duality with linear monopole and anti-monopole superpotential 
with $f+2$ flavors
and assigning two large and opposite real masses  to two pairs of fundamentals and anti-fundamentals.

In this case we have more freedom than above in the choice of the masses, and this gives rise to a free FI in both the electric and magnetic side. Furthermore on the magnetic side we have to consider the Higgsing to 
$U(3kf-n) \times U(k-1)^2$. In the first sector there are $f$ fundamental flavors while in the other two sectors there is a single fundamental flavor. Each sector contains two adjoints interacting through the usual power law binomial superpotential
and there are interactions between the light singlets surviving the real mass flow and coming from the original meson and
other combinations of fundamental and adjoint fields, generalizing the construction in (\ref{Wsingle0}).
There linear monopole superpotentials for the $U(\nn)$ and the $U(3kf-n)$ sector are lifted, while there is still a single linear monopole superpotential in each $U(k-1)$ sector. Furthermore the dressed monopole and the anti-monopoles of the 
$U(3kf-n)$ interact through an AHW-like superpotential with the ones of the two $U(k-1)$ sectors (with the sign choice given by the sign choice of the large mass limit in the electric side).
Proceeding as above we can dualize both the $U(k-1)$ sectors and we end up with the duality of \cite{Hwang:2018uyj}.

This construction can be reproduced on the partition function as well. 
The real masses are chosen as
\begin{equation}
\left\{
\begin{array}{ccc}
m_{f+1} &=& \mathbf{m}+\frac{\xi}{2} +s
\\
m_{f+2} &=& \mathbf{m} -\frac{\xi}{2} +s
\\
n_{f+1} &=& \mathbf{m} +\frac{\xi}{2} -s
\\
n_{f+1} &=& \mathbf{m} -\frac{\xi}{2}-s
\end{array}
\right.,
\end{equation}

On the dual side the gauge group $U(3kf-n+2k-2)$ is Higgsed to 
$U(3kf-n) \times U(k-1) \times U(k-1)$ by shifting $(k-1)$ integration variables as $\sigma_i\rightarrow \sigma_i-s$
and  $(k-1)$ integration variables as $\sigma_i\rightarrow \sigma_i+s$.
Computing the large $s$ limit on the identity (\ref{firstU}) and eliminating the divergent part, that coincides on the electric and on magnetic side we end up with the relation

\begin{eqnarray}
\label{firstU2}
Z_{U(\nn)}^{f} (\vec m; \vec n,;\tau_A,\tau_B;\xi)
&=&
Z_{U(3kf-n)}^{f} (\tau_A-\tau_B-\vec n; \tau_A-\tau_B-\vec n ;\tau_A,\tau_B;-\xi)
\nonumber \\
&\times&
\prod_{j=0}^{k-1}\prod_{\ell=0}^2 \prod_{a,b=1}^{f}  \Gamma_h(j \tau_A+\ell \tau_B+m_a+n_b)
\nonumber \\
&\times&
\prod_{\eta=\pm1}
Z_{U(k-1)}^1 (\widetilde{\mathbf{m}}_\eta,\widetilde{\mathbf{m}}_\eta,\tau_A,\tau_B; 2\eta(\mathbf{m}- \omega) -\xi)
\nonumber \\
\times&&
\prod_{j=0}^{k-1}\prod_{\ell=0}^2 
\prod_{a,b=1}^{f}  \Gamma_h(j \tau_A+\ell \tau_B+m_a+n_b) 
\nonumber \\
\times&&
\prod_{j=0}^{k-1}\prod_{\ell=0}^2 \prod_{a,b=1}^{f} \prod_{\eta=\pm1} \Gamma_h(j \tau_A+\ell \tau_B+(2m+\eta \xi))
\nonumber \\
\end{eqnarray}
with $\widetilde{\mathbf{m}}_\eta = \tau_A-\tau_B-\mathbf{m} +  \frac{\eta \xi}{2}$.
The constraint on the parameters reads
 \begin{equation}
 \sum _{a=1}^f (m_a+n_a)=2 (f \omega -(n-1) (\tau _A+\tau _B-\omega ))-4 \mathbf{m}
 \end{equation}
In this case it signals the presence of the AHW interactions and it 
 plays an important role in identifying the monopole operators of the confining sectors with the ones of
 the electric theory.
 On the other hand it does not play any role on the $U(\nn)$ and $U(3kf-n)$ sectors because we are in 
 presence of an unconstrained FI $\xi$. It signals the absence of linear monopole deformations in 
 these two gauge sectors.

Proceeding as above by dualizing the two $U(k-1)$ sectors with the help of (\ref{conj}) and eliminating
 the massive fields with the help of the inversion formula for the hyperbolic Gamma functions we end up with the relation
\begin{eqnarray}
\label{lastU}
Z_{U(\nn)}^{f} (\vec m; \vec n,;\tau_A,\tau_B;\xi)
&=&
Z_{U(3kf-n)}^{f} (\tau_A-\tau_B-\vec n; \tau_A-\tau_B-\vec n ;\tau_A,\tau_B;-\xi)
\nonumber \\
&\times&
\prod_{j=0}^{k-1}\prod_{\ell=0}^2 \prod_{a,b=1}^{f}  \Gamma_h(j \tau_A+\ell \tau_B+m_a+n_b)
\nonumber \\
&\times&
\prod_{q=0}^{\frac{k-3}{2}}
\Gamma_h(\pm 2\xi + 2 f \omega \!-\!(n-1) \tau _A \!-\! \sum _{a=1}^f (m _a+n _a)\!+\!(2 q+1) \tau_A)
\nonumber \\
&\times&\!\!\!\! \!\!\!\!\!\!
\prod_{\tiny{
\begin{array}{c}
j=0,\dots,k-1\\
\ell=0,\dots,2\\
j \ell=0
\end{array}
}}
\!\!\!\!\!\! \!\!\!\!
\Gamma_h(\pm \xi +  f \omega-\frac{n-1}{2} \tau _A -\frac{1}{2} \sum _{a=1}^f (m _a+n _a)+\ell \tau_B+j \tau_A)
\nonumber \\
\end{eqnarray}
with unconstrained parameters. 
More precisely the parameters $m_a$ and $n_a$ are constrained as 
$\sum_a m_a=\sum_a n_a = f m_A$ where $m_A$ is the parameter associated to the axial symmetry.
The relation (\ref{lastU}) is the expected one between the electric and the magnetic side of the duality 
found in \cite{Hwang:2018uyj}.

%
%
%
%
\subsection{$Usp(2\nn)$ without  monopole superpotential}
 \label{subsec:noTUSp}
%
%
%
%
Another duality that can be derived from the reduction of $Usp(2\nn)$ on the circle 
is obtained by eliminating the superpotentials $\eta  Y_{USp}$ and $\tilde \eta y_{Usp}$ in 
(\ref{Worch}) and in (\ref{Worchdual}).
The flow in this case is engineered by considering $2(f+1)$ fundamentals and by assigning a large and opposite mass to
two of them.
On the dual side the gauge group is Higgsed to $USp(3k-n-2k-1) \times U(k-1)$, with $2f$ light fundamentals $q$ in the 
symplectic sector and one fundamental flavor $(p,\tilde p)$ in the $U(k-1)$ sector.
Again the FI term generated in the $U(k-1)$ sector is constrained and it signals the presence of a linear monopole or anti-monopole 
superpotential (here the choice depends on the sign of the shift in the real scalar in the vector multiplet).
In the $USp(3k-n-2k-1)$ sector there are two anti-symmetric rank-two tensors $a$ and $b$ while in the 
$U(k-1)$ sector there are two adjoints $x$ and $y$.
The superpotential of this dual model is
\begin{eqnarray}
W &=& Tr a^{k+1} + Tr a b^2 + Tr x^{k+1}  + Tr x y^2 + V_{0,0}^+
+ \nonumber \\
&+&
\sum_{j=0}^{k-1} \sum_{\ell=0}^{2} (M_{j,\ell} q a^j b^\ell q  +N_{j,\ell} p x^j y^\ell  \tilde p)
+
\nonumber \\
&+& \sum_{j=1}^{k-1} V_{j,0}^- \widetilde Y_{k-j,0}
+\sum_{\ell=0}^{2} V_{0,\ell}^- \widetilde Y_{0,2-\ell}
+ \sum_{q=0}^{\kappa=\frac{k-3}{2}}  W_{q}^- \widetilde Z_{\kappa-q}
\end{eqnarray}
where $\widetilde Y_{j,\ell}$ and  $\widetilde Z_{q}$ are  dressed monopoles of the $USp(3kf-n-2k-1)$ sector
with topological charge $1$ and $2$ respectively.
They can be defined by radial quantization as in (\ref{mondef}).
Dualizing the $U(k-1)$ sector we arrive at the final formulation of the duality, where the superpotential becomes
\begin{eqnarray}
\label{finalUSpAha}
W &=& Tr a^{k+1} + Tr a b^2 + \sum_{j=0}^{k-1} \sum_{\ell=0}^{2} (M_{j,\ell} q a^j b^\ell q )
+
\nonumber \\
&+& \sum_{j=1}^{k-1} Y_{j,0} \widetilde Y_{k-j,0}
+\sum_{\ell=0}^{2} Y_{0,\ell} \widetilde Y_{0,2-\ell}
+ \sum_{q=0}^{\kappa=\frac{k-3}{2}}  Y_{q} \widetilde Z_{\kappa-q}
\end{eqnarray}
where the monopoles arising from the $U(k-1)$ sectors are identified with the 
ones of the electric $Usp(2\nn)$ gauge theory.

The global symmetry in this case is $SU(2f) \times U(1)_A \times U(1)_R $ and 
the fields and the monopoles transform under the gauge and global symmetries as follows:
\begin{equation}
\begin{array}{|c|c|c|c|c|c|}
\hline
\text{Field} & Usp(2\nn) & USp(2 \tilde n) & SU(2f) &U(1)_A& U(1)_R \\
\hline
Q &n&1&2f& 1& r \\
\hline
A &n(2n\!-\!1)&1&1&0&\frac{2}{k+1} \\ 
\hline
B &n(2n\!-\!1)&1&1&0&\frac{k}{k+1} \\
\hline
Y_{j \ell} & 1 & 1 & 1 & -2f & 2f(1-r) + \frac{2 j + k \ell-2(n+k)}{k+1}\\
\hline
Z_q& 1 & 1 & 1 & -4f &4f(1-r) + \frac{2+4q-4(n+k)}{k+1}\\
\hline
q&1&\tilde n&\overline{2f}&-1& \frac{2-k}{k+1} - r \\
\hline
a&1&\tilde n(2\tilde n\!-\!1)&0&&\frac{2}{k+1} \\ 
\hline
b&1&\tilde n(2\tilde n\!-\!1)&1&0&\frac{k}{k+1} \\ 
\hline
{\mathcal{M}_{j \,0}}_{(j=0,\dots,k-1)}&1&1& f(2f\!-\!1)&0& 2(1\!-\!r)+\frac{2j}{k+1} \\
\hline
{\mathcal{M}_{2j\, 1}}_{(j=0,\dots,\frac{k-1}{2})}&1&1&  f(2f\!-\!1)&2&  2(1\!-\!r)+\frac{4j+k}{k+1}\\
\hline
{\mathcal{M}_{2j+1\, 1}}_{(j=0,\dots,\frac{k-3}{2})}&1&1& f(2f\!+\!1) &2&2(1\!-\!r)+\frac{4j+k+2}{k+1} \\
\hline
{\mathcal{M}_{j \,2}}_{ (j=0,\dots,k-1)}&1&1&  f(2f\!-\!1)& 2&2(1-r)+\frac{2j+2k}{k+1} \\
\hline
\rule{0pt}{13pt} 
{\widetilde Y}_{j \ell}& 1 & 1 & 1 & 2f & 2(f\!-\!1)r + \frac{k\ell + 2(j+n+k+1)}{k+1} \\
\hline
\rule{0pt}{13pt} 
{\widetilde Z}_q& 1 & 1 & 1 & 4f & 4f (r\!-\!1)+\frac{4(q+n+k)+6}{k+1}
\\
\hline
\end{array}
\end{equation}

We conclude the discussion by performing such a flow on the partition function.
In this case we consider $2f+2$ mass parameters $\mu_a$ and assign the masses as
\begin{equation}
\left\{
\begin{array}{ccc}
\mu_{f+1} &=& \mathbf{m} +s
\\
\mu_{f+2} &=& \mathbf{m} -s
\end{array}
\right.,
\end{equation}
The dual Higgsing corresponds to the shift $\sigma_i \rightarrow \sigma_i+s$ 
(or equivalently $\sigma_i \rightarrow \sigma_i+s$) for $k-1$ integration variables.
Computing the large $s$ limit on the identity (\ref{firstidusp}) 
and eliminating the divergent part, 
that coincides on the electric and on magnetic side, we end up with the relation
\begin{eqnarray}
\label{firstidusp2}
Z_{Usp(2\nn)}^{2 f} (\vec \mu;\tau_A,\tau_B)
&=&
Z_{USp(2(3kf-n-2k-1))}^{2 f} (\tau_A-\tau_B-\vec \mu;\tau_A,\tau_B)
\nonumber \\
&\times&
\prod_{j=0}^{k-1}\prod_{\ell=0}^2 \prod_{1\leq a<b\leq f}  \Gamma_h(j \tau_A+\ell \tau_B+\mu_a+\mu_b)
\nonumber \\
&\times&
\prod_{q=0}^{k-1}\prod_{a=1}^f \Gamma_h((2q+1) \tau_A+\tau_B+2\mu_a) 
\nonumber \\
&\times&
Z_{U(k-1)}^{1} (\tilde {\mathbf{m}} , \tilde {\mathbf{m}} ;\tau_A,\tau_B,-2(\omega-\mathbf{m}))
\nonumber \\
&\times&
\prod_{j=0}^{k-1}\prod_{\ell=0}^2 \prod_{1\leq a<b\leq f}  \Gamma_h(j \tau_A+\ell \tau_B+2\mathbf{m})
\nonumber \\
\end{eqnarray}
with $\tilde {\mathbf{m}} =  \tau_A-\tau_B-\mathbf{m}$ and the constraint
\begin{equation}
\label{bcfin}
\sum_{a=1}^{2 f} \mu_a = 2(\omega(f-1)-(n-1) (\tau_A+\tau_B-\omega)) -2 \mathbf{m}
\end{equation}
Again this constraints does not affect the mass parameters on the electric side, signaling that the KK monopole
superpotential has been lifted. On the dual side the constraint is crucial for the identification of the monopoles (or anti-monopoles) of the confining $U(k-1)$ sector with the ones of the electric $Usp(2\nn)$ model.
The last step consists of integrating  $Z_{U(k-1)}^1$ using the relation (\ref{conj}) and arrive to the equality 
\begin{eqnarray}
\label{firstidusp3}
Z_{Usp(2\nn)}^{2 f} (\vec \mu \! \!&\! \!;\!\!&\! \!\tau_A,\tau_B)
=
Z_{USp(2(3kf-n-2k-1))}^{2 f} (\tau_A-\tau_B-\vec \mu;\tau_A,\tau_B)
\nonumber \\
&\times&
\prod_{j=0}^{k-1}\prod_{\ell=0}^2 \prod_{1\leq a<b\leq f}  \Gamma_h(j \tau_A\!+\!\ell \tau_B\!+\!\mu_a\!+\!\mu_b)
\prod_{q=0}^{\frac{k-3}{2}}\prod_{a=1}^f \Gamma_h((2q+1) \tau_A\!+\!\tau_B\!+\!2\mu_a) 
\nonumber \\
&\times&\!\!\!\! \!\!\!\!\!\!
\prod_{\tiny{
\begin{array}{c}
j=0,\dots,k-1\\
\ell=0,\dots,2\\
j \ell=0
\end{array}
}}
\!\!\!\!\!\! \!\!\!\! \Gamma_h(j\tau_A+\ell \tau_B+2 f \omega \!-\! \tau_A (n+k) \!-\!\sum _{a=1}^{2 f} \mu _a)
\nonumber \\
&\times&
\prod_{q=0}^{\frac{k-3}{2}}  \Gamma_h((2q+1) \tau_A +4 f \omega \!-\! 2\tau_A (n+k)\!-\!2\sum _{a=1}^{2 f} \mu _a)
\end{eqnarray}
%
%
%
%
\section{Reconsidering the 4d/3d reduction of $U(\nn)$ SQCD with two adjoints}
\label{sec:brodie43}
%
%
%
%

In this section we study the reduction of the 4d duality worked out in \cite{Brodie:1996vx}
and reviewed in sub-section \ref{subsec:4d} along the lines of \cite{Aharony:2013dha}.

Some comments are in order. First we consider a slight modification of the original the duality, by gauging the baryonic 
symmetry. The $SU(\nn)$ case can be studied similarly, even if it modifies the structure of the effective superpotential 
generated in the circle reduction. It is nevertheless possible to recover the 3d limit of the $SU(\nn)$ duality by gauging the 
topological symmetry in the final step of our procedure.
We will further comment on this possibility in the conclusions.
A second comment regards the restriction to the case of odd $k$. A similar duality was also conjectured for the case of  even $k$ in \cite{Brodie:1996vx}. The difference in such case is that the chiral ring is not truncated at classical level by the F-term
constraints on the electric side and the proposal is that such truncation appears quantum mechanically. Here we will not elaborate further on this case and we restrict to the odd case.

The reduction of this duality to an  \emph{effective} duality on the circle has been studied in \cite{Hwang:2018uyj} by breaking the gauge group, and by reducing the duality only in $U(n_i)$ sectors without adjoints. 
Here we adopt a different strategy and leave the gauge group unbroken. The motivation underlining our analysis is that we want to follow the various steps at the level of the reduction of the 4d superconformal index to the 3d  squashed three sphere partition function.

On the circle we obtain an \emph{effective} duality relating
\begin{itemize}
\item 3d $U(\nn)$ SQCD with $f$ flavors $Q$ and $\widetilde Q$ with two adjoints $X$ and $Y$ 
interacting through the superpotential $W = (\ref{Brodieel}) + W_{\eta}$
with
\begin{equation}
\label{etael}
W=Tr X^{k+1} + Tr X Y^2+\eta \bigg(\sum_{j=1}^{k-1}V_{j,0}^+ V_{k-j,0}^-+\sum_{l=0}^{2}V_{0,l}^+V_{0,2-l}^{-}+\sum_{q=0}^{\kappa=\frac{k-3}{2}}W_{q}^+W_{\kappa-q}^{-} \bigg)
\end{equation}
\item 3d $U(3kf-n)$ SQCD with $f$ dual flavors $q$ and $\widetilde q$ 
 with two adjoints $x$ and $y$ 
interacting through the superpotential  $W = (\ref{Brodiemag}) + W_{\widetilde \eta}$
\begin{equation}
\label{etamag}
W_{\widetilde \eta} =\widetilde\eta \bigg( \sum_{j=1}^{k-1}\widetilde V_{j,0}^+ \widetilde V_{k-j,0}^-+\sum_{l=0}^{2}\widetilde V_{0,l}^+\widetilde V_{0,2-l}^{-}+\sum_{q=0}^{\kappa=\frac{k-3}{2}}\widetilde W_{q}^+\widetilde W_{\kappa-q}^{-}\bigg)
\end{equation}
\end{itemize}
The superpotentials $W_\eta$ and $W_{\widetilde \eta}$ break the axial symmetry, anomalous in the 4d case. 
The conventional 3d duality is obtained by a real mass flow, accompanied on the magnetic side by a dual Higgsing of the gauge group. 
Such real mass flow can be engineered on the electric side by considering $f+2$ flavors and by assigning large (and opposite) real mass to two pairs of  fundamentals and  anti-fundamentals. 
On the magnetic side we can assign the same opposite shifts to the scalars in the vector multiplet by breaking 
$U(3k(f+2) -n ) \rightarrow U(3kf-n) \times U(3k)^2$.
In the $U(3kf-n)$ gauge sector we have two adjoint $x$ and $y$ and $f$ pairs of fundamental and anti--fundamental $q$ and $\widetilde q$. In the $U(3k)$ sectors we still have two adjoints, denoted respectively as $x_\rho$ and $y_\rho$ in the first sector and $x_\xi$ and $y_{\xi}$ in the second sector and one flavor, denoted $(q_{\rho},\widetilde q_\rho)$ and $(q_{\xi},\widetilde q_{\xi})$.
The superpotential is 
\begin{eqnarray}
\label{Wdualafterrm}
W &=& Tr x y^2+ Tr x^{k+1} + Tr x_\rho y_\rho^2+ Tr x_\rho^{k+1} +Tr x_\xi y_\xi^2+ Tr x_{\xi}^{k+1} 
\nonumber \\
&+&\sum_{j=0}^{k-1} \sum_{\ell=0}^{2} \Tr (\mathcal{M}^{j,\ell} q x^{k-1-j}y^{2-\ell} \widetilde q  +\mathcal{M}_\rho^{j,\ell} q_{\rho} x_\rho^{k-1-j} y_\rho^{2-\ell} \widetilde q_{\rho} +
\mathcal{M}_\xi^{j,\ell} q_{\xi} x_\xi^{k-1-j} y_\xi^{2-\ell} \widetilde q_{\xi} )
\nonumber \\
&+&   \sum_{j=1}^{k-1}(\widetilde V_{j,0}^+ \widetilde {\mathbf{V}}_{k-j,0}^{-}+\widetilde V_{j,0}^- \widetilde {\mathcal{V}}_{k-j,0}^{+}+\widetilde {\mathbf{V}}_{j,0}^+ \widetilde {\mathcal{V}}_{k-j,0}^{-})
\nonumber \\
&+&
\sum_{\ell=0}^{2}(\widetilde V_{0,\ell}^+ \widetilde {\mathbf{V}}_{0,2-\ell}^{-}+\widetilde V_{0,\ell}^- \widetilde {\mathcal{V}}_{0,2-\ell}^{+}+\widetilde {\mathbf{V}}_{0,\ell}^+ \widetilde {\mathcal{V}}_{0,2-\ell}^{-})
\nonumber \\
&+&
\sum_{q=0}^{\kappa=\frac{k-3}{2}}
(\widetilde W_{q}^+\widetilde {\mathbf{W}}_{\kappa-q}^{-}+
\widetilde W_{q}^-\widetilde {\mathcal{W}}_{\kappa-q}^{+}
+
\widetilde {\mathbf{W}}_{q}^+ \widetilde {\mathcal{W}}_{\kappa-q}^{-})
\end{eqnarray}
The first two lines of (\ref{Wdualafterrm}) correspond to the superpotential (\ref{Brodiemag}) in the gauge sectors
of the dual theory, inherited from the original one after the real mass flow and the Higgs flow.
The last three lines on the other hand correspond to an AHW-like superpotential generated between the monopole operators because of the dual Higgsing.
We have denoted as $\mathbf{V}$ and $\mathbf{W}$ the monopoles of $U(3k)_{\rho}$ sector and 
$\mathcal{V}$ and $\mathcal{W}$ the monopoles of $U(3k)_{\xi}$ sector.

The last step consists of dualizing the $U(3k)$ sectors in terms of singlets. The duality that we have to use in this case
corresponds to the limiting case of the 3d duality that we are looking for.
This is a very standard phenomenon in the reduction of 4d dualities to 3d. 
In the case of SQCD such a limiting case was interpreted independently  as local mirror symmetry, indeed the extra sectors correspond to SQED with one flavor, and the confining model could be regarded as local mirror symmetry. In the case with one adjoint such sector correspond to $U(k)$ with one adjoint and one flavor and $W=Tr X^{k+1}$. This model can be shown to be confining as well, even in absence of the adjoint superpotential, as discussed for example in \cite{Aghaei:2017xqe,Nieri:2018pev}.
In both cases the confining duality corresponds to the limiting case of the duality that one is looking for.

Here we borrow the results obtained in sub-section \ref{subsec:noTU} and dualize the two 
$U(3k)$ sectors with a fundamental flavor.
From the superpotential (\ref{Wdualafterrm}) one can observe that the mesons $q_{\xi} x_\xi^{k-1-j} y_\xi^{2-\ell} \widetilde q_{\xi} $ and 
$q_{\rho} x_\rho^{k-1-j} y_\rho^{2-\ell} \widetilde q_{\rho} $
become massive and they can be integrated out at zero vev.
In a similar fashion the monopoles 
$\widetilde {\mathbf{V}}_{j,\ell}^+ $ and $\widetilde {\mathcal{V}}_{j,\ell}^{-}$ with $j \ell=0$
and  $\widetilde {\mathbf{W}}_{q}^+$ and $\widetilde {\mathcal{W}}_{q}^{-}$ with $q=0,\dots,\frac{k-3}{2}$
are singlets after confining the $U(3k)$ gauge groups and they become  massive because of the 
superpotential  (\ref{Wdualafterrm}).
On the other hand the monopoles 
$\widetilde {\mathbf{V}}_{j,\ell}^- $ and $\widetilde {\mathcal{V}}_{j,\ell}^{+}$ with $j \ell=0$
and  $\widetilde {\mathbf{W}}_{q}^-$ and $\widetilde {\mathcal{W}}_{q}^{+}$ with $q=0,\dots,\frac{k-3}{2}$
are singlets of the $U(3 f-n)$ gauge theory that interact with the monopoles
$\widetilde {V}_{j,\ell}^\pm $ and  $\widetilde {W}_{q}^{\pm}$.
They can then be naturally identified with the monopoles 
 $ {V}_{j,\ell}^\pm $ and  $ {W}_{q}^{\pm}$ of the electric theory acting as singlets in the dual phase.
 All in all we arrive at the dual superpotential (\ref{Wdualfinal})
 corresponding as expected to the one obtained in \cite{Hwang:2018uyj}.

 This analysis can be reproduced on the partition function as well. The first step consists of 
 reducing the identity of \cite{Spiridonov:2009za} between the 4d indices using the prescription of \cite{Aharony:2013dha}.
 This gives an identity reproducing the effective duality on $S^1$, with a constraint between the mass parameters
 signaling the presence of the KK monopole superpotential.
 The identity for the effective duality is
 \begin{eqnarray}
 \label{brodieS1}
 Z_{U(\nn)}^{f} (\vec m, \vec n;\tau_X,\tau_Y;\xi) 
 &=&
  Z_{U(3kf-n)}^{f} (\vec {\widetilde n}, \vec {\widetilde n};\tau_X,\tau_Y;-\xi) 
  \nonumber \\
&\times& 
\prod_{j=0}^{k-1}\prod_{\ell=0}^2 
\prod_{a,b=1}^{f}  \Gamma_h(j \tau_X+\ell \tau_Y+m_a+n_b) 
 \end{eqnarray}
 where $\widetilde m_a = \tau_X-\tau_Y-m_a$ and $\widetilde n_a = \tau_X-\tau_Y-n_a$.
 In this case the FI parameter is free while
 the constraint between the mass parameters corresponds to (\ref{bcs1}),
 preventing the generation of an axial symmetry.

Next we need to engineer the flow to the pure 3d duality. This flow can be constructed by considering $f+2$ flavors and  parameterizing the last two pairs as
\begin{equation}
\left\{
\begin{array}{ccc}
m_{f+1} &=& \mathbf{m} +s
\\
m_{f+2} &=& \mathbf{m} -s
\\
n_{f+1} &=& \mathbf{m}  -s
\\
n_{f+1} &=& \mathbf{m} +s
\end{array}
\right.,
\end{equation}

On the dual side the gauge group $U(3k(f+2)-n)$ is Higgsed to 
$U(3kf-n) \times U(3k) \times U(3k)$, by shifting $3k$ integration variables as $\sigma_i\rightarrow \sigma_i-s$
and  $3k$ integration variables as $\sigma_i\rightarrow \sigma_i+s$.
Computing the large $s$ limit on the identity (\ref{firstU}) and eliminating the divergent part, that coincides on the electric and on magnetic side we end up with the relation
  
\begin{eqnarray}
 \label{brodie3d3ksectors}
 Z_{U(\nn)}^{f} (\vec m, \vec n;\tau_X,\tau_Y;\xi) 
 &=&
  Z_{U(3kf-n)}^{f} (\vec {\widetilde n}, \vec {\widetilde n};\tau_X,\tau_Y;-\xi) 
   \nonumber \\
&\times& 
\prod_{\eta=\pm 1}  Z_{U(3k)}^{1} (\widetilde{\mathbf{m}},  ;\tau_X,\tau_Y;2 \eta(\mathbf{m}-\omega) -\xi)
 \\
&\times& 
\prod_{j=0}^{k-1}\prod_{\ell=0}^2 
\prod_{a,b=1}^{f}  \Gamma_h(j \tau_X+\ell \tau_Y+m_a+n_b)  \Gamma_h(j \tau_X+\ell \tau_Y+2 \mathbf{m})^2
 \nonumber
 \end{eqnarray}
where $\widetilde{\mathbf{m}} = \tau_X-\tau_Y-\mathbf{m}$.
The $U(3k)$ sectors are confining and the corresponding integrals can be computed using the results of the previous section.   Indeed they correspond to the limiting case of the identity (\ref{lastU}), because the rank of the dual gauge group vanishes if $n=3k$ and $f=1$. By computing these integrals, simplifying the massive fields and using the  constraint
 \begin{equation}
 \sum_{a=1}^{f} (m_a+n_a) +4 \mathbf{m} = 2 \omega (f+2)-n \tau_X
 \end{equation}
 we arrive at the final relation, that coincides with (\ref{lastU}) as expected.

%
%
%
\section{Conclusions}
\label{sec:conc}
%
%
%
%

In this paper we have studied 3d $\mathcal{N}=2$ dualities for $U(\nn)$ and $Usp(2\nn)$ SQCD 
with two adjoints and two antisymmetric rank-two tensors respectively.
These dualities descend from the 4d parents worked out in \cite{Brodie:1996vx} and \cite{Brodie:1996xm}.
By real mass and Higgs flows we have constructed many other dualities recovering some results  
obtained in \cite{Hwang:2018uyj} and \cite{Hwang:2022jjs} and finding new pairs as well.
Our results are based on a conjectured confining duality discussed in sub-section \ref{subsec:uTp}
for $U(k-1)$ SQCD with two adjoint and a linear monopole superpotential.
By using this conjectured duality we have  constructed a consistent picture, mapping the various dualities,
extending the cases without adjoints and with one adjoint in a uniform manner.
Furthermore we have checked the consistency of the various flows by the analysis of the three sphere partition
function. In this case  the conjectured confining $U(k-1)$  duality corresponds to assume the relation (\ref{conj}).

Many open questions arise from this work.
First the prove of formula (\ref{conj}) is a necessary step to confirm the validity of our approach
and corroborate the claims that we made for the new dualities. 
Another open question is related to the other duality
\footnote{Observe that in \cite{Hwang:2022jjs} many other dualities are conjectured in presence of linear monopole deformations for the dressed  monopole operators, in the case with a single adjoint. Similar results are then expected for the 
case with two adjoints. Here we have not addressed such an issue from the 4d perspective.} 
discussed in \cite{Hwang:2022jjs}, and reviewed in sub-section \ref{subsec:3d}, involving linear  superpotentials for the  monopole operators with charge 2 under $U(1)_T$. It would be interesting to obtain such a duality by flowing from the  
ones discussed here.

Further classes of dualities can then be constructed starting from the ones proposed in our work, involving 
CS terms,  a \emph{chiral} matter content (i.e. a different number of fundamentals and anti-fundamentals)
and/or $SU(\nn)$ gauge groups. 
The 3d analog of Brodie duality in presence of CS terms has been already proposed in \cite{Niarchos:2009aa}. 
It would be interesting to obtain this last duality by flowing from the ones without CS terms and then generalize the construction to the $Usp(2\nn)$ case as well.
Also the cases with a chiral matter content can in principle be studied by opportune real mass flows, 
with possible dual Higgs flow, along the lines of \cite{Benini:2011mf,Aharony:2014uya,Hwang:2015wna,Amariti:2020xqm}.
The analysis of $SU(\nn)$ dualities is in general less straightforward. This is due to the presence of quadratic monopole operators
in many of the dual phases. 
The general prescription to obtain an  $SU(\nn)$ gauge group  starting from 
$U(\nn)$ consists of 
promoting the topological symmetry to a gauge symmetry; in presence of an FI term this gauging 
leads to a mass term   between  the dynamical photon of $U(1)_T$ 
and the original photon of $U(1) \subset U(\nn)$, through a mixed CS term. By integrating out the massive fields one is left, on the electric side, with an $SU(\nn)$ gauge group. On the other hand the presence of monopole operators in the dual phase  does not allow the same operation 
in a straightforward manner, because there are extra fields  charged under the gauged topological symmetry. 
The interpretation of the charge two monopole operators arising from the gauging of $U(1)_{T}$ is  
less clear than the one for the monopole with charge one under $U(1)_T$ and it is not obvious how to manage this new gauged sector in general.
Anyway one could reverse the logic and start from the dualities that we have obtained here before applying any local confining duality on the extra  $U(k-1)$ sectors. We hope to come back to this problem in the future.
Other interesting lines of research involve the generalization of the construction to the other dualities with two 
rank-two tensors proposed in \cite{Brodie:1996xm}.

Another necessary integration of our results  is the 
analysis of the relevancy of the monopole deformations. This can be studied by F-maximization
\cite{Jafferis:2010un,Jafferis:2011zi,Closset:2012vg}, generalizing the analysis of \cite{Benini:2017dud} for the case of
SQCD. 
More generally F-maximization would be necessary to analyze the conformal windows of the models 
treated here. A similar study for 4d models with two adjoints was performed in \cite{Intriligator:2003mi,Okuda:2005me,Parnachev:2008yt} 
by using  a-maximization. 

The interpretation of the dimensional reduction of the 4d dualities in terms of  T-duality on the corresponding  brane picture, along the scheme proposed in \cite{Amariti:2015yea,Amariti:2015mva,Amariti:2016kat}, does not look possible here, because the Hanany-Witten setup engineering  the 4d models is missing. Anyway a possible 4d brane interpretation 
has been furnished in \cite{Ahn:1997yc}, generalizing the approach of \cite{Ooguri:1997ih} to che case of Brodie duality.
It would be then interesting to investigate  the geometric reduction of the 4d $U(\nn)$ duality with two adjoints in this way.

 As a last comment we would like to stress that all the models discussed here refer to the case of odd $k$, being $k+1$ the exponent of the adjoint $X$ and of the anti-symmetric $A$ in all the $U(\nn)$ and $Usp(2\nn)$ SQCD superpotentials respectively.
While this choice is motivated by anomaly matching for the $Usp(2\nn)$ gauge groups, $U(\nn)$ gauge theories
with  $k$-even  are more tricky, because the chiral ring is supposed to truncate at quantum level, differently from the classical truncation taking place for $k$-odd  (see \cite{Mazzucato:2005fe,Intriligator:2016sgx,Bajc:2019vbp}
  for further discussions on this issue). 
In three dimensions we have seen that the $U(\nn)$ dualities can be derived from the $Usp(2\nn)$ one, derived by circle reduction, only for odd $k$. 
 It would be then interesting to understand if is there any connection between 
 the quantum truncation of the chiral ring and the  4d/3d reduction.

%
%
%
%
%
%
\section*{Acknowledgments}
%
%
This work has been supported in part by the Italian Ministero dell'Istruzione, 
Universit\`a e Ricerca (MIUR), in part by Istituto Nazionale di Fisica Nucleare (INFN) through the “Gauge Theories, Strings, Supergravity” (GSS) research project and in part by MIUR-PRIN contract 2017CC72MK-003.  

\appendix
%
%
\section{Flow from $Usp(2n)$ without monopole superpotential to $U(n)$ with $W_{mon}= V^+_{0,0}$}
\label{sec:A}
%
%
In this section we study a flow connecting the pure 3d duality for symplectic gauge group  \ref{subsec:noTUSp} and the duality for $U(n)$ gauge group with a single monopole (or anti-monopole) superpotential \ref{subsec:uTp}. This provides a consistency check of the web of dualities summarized in Figure \ref{ease}.

The flow is triggered by giving large positive mass to half of the fundamentals of the $Usp(2n)$ theory and opposite mass to the other fundamentals, together with an Higgsing of all the scalars in the vector multiplets both in the electric and in the magnetic theory. When all the massive fields are integrated out one obtains a $U(n)$ gauge theory with two adjoints, $f$ flavors and a linear monopole superpotential $W_{mon} = V^+_{0,0}$. One can also obtain the theory with linear anti-monopole superpotential $W_{mon} = V^-_{0,0}$. The sign depends on the choice of the sign of the v.e.v. of the scalars in the vector multiplets.

The flavor symmetry $SU(2f) \times U(1)_A \times U(1)_R$ is broken to $SU(f)_L \times SU(f)_R \times U(1)_{\tilde{A}} \times U(1)_R$, where $U(1)_{\tilde{A}}$ is the combination of the axial and topological symmetry that is preserved by the monopole superpotential.

We can follow this flow on the partition function as follows. We start from the identity \eqref{firstidusp3} and assign the masses as: 
\begin{equation}
\label{eq:masses_Usp_U}
\left\{
\begin{array}{ccc}
\mu_{a} &=& m_a +s
\\
\mu_{f+a} &=& n_a -s
\end{array}
\right. \quad a = 1,\dots,f
\end{equation}
The Higgsing corresponds to the shift $\sigma_i \to \sigma_i + s$ for $i=1,\dots, n$. Computing the large $s$ limit and eliminating the divergent part we end up with the relation \eqref{finalT}, which corresponds to the duality for $U(n)$ with linear monopole superpotential.\\

\subsection{The cases with one adjoint and without adjoints}

Analogous flows can be studied in the case of Aharony duality and in the presence of a single antisymmetric tensor with $A_k$-type superpotential. These flows connect a duality with symplectic gauge groups to a duality with unitary gauge group and linear monopole superpotential.
As far as we were able to check these flows have not been discussed in the literature.

These flows are performed in a similar fashion as the one discussed above for the case of two antisymmetric tensors. 
In the case of Aharony duality we start from the duality for $Usp(2n)$ gauge group \cite{Willett:2011gp} with $2f$ fundamentals and give large positive mass to $f$ fundamentals and negative mass to the other $f$ fundamentals together with an Higgsing of all the scalars in the vector multiplets. 
Integrating out the massive fields we end up with the duality for $U(n)$ gauge group with $f$ flavors and linear monopole superpotential, originally studied in \cite{Benini:2017dud}. The flavor symmetry is broken in the same way as in the case of two antisymmetric tensors discussed above. 

Analogously we can start from the duality for $Usp(2n)$ gauge theory with $2f$ fundamentals, one antisymmetric and $A_k$-type superpotential \cite{Intriligator:1995ax}. By giving large masses and Higgsing the gauge group in the same way as the cases discussed above we end up with the duality for $U(n)$ gauge group with $f$ flavors, one adjoint with $A_k$-type superpotential and linear monopole superpotential \cite{Amariti:2018wht}. 

We can follow these flows on the partition function by giving large masses as in \eqref{eq:masses_Usp_U}, together with the shift $\sigma_i \to \sigma_i + s$. In the large $s$ limit the divergences cancel between the dual theories and we end up with the identities corresponding to the dualities for $U(n)$ with a single linear monopole superpotential obtained in \cite{Benini:2017dud} and in \cite{Amariti:2018wht}.

\bibliographystyle{JHEP}
\bibliography{ref}

\end{document}